\documentclass[aps,pre,superscriptaddress,twocolumn,showpacs,nofootinbib,amsmath]{revtex4}
\usepackage{graphicx}

\newcommand{\be}{\begin{equation}}
\newcommand{\ee}{\end{equation}}
\newcommand{\bse}{\begin{subequations}}
\newcommand{\ese}{\end{subequations}}
\newcommand{\bea}{\begin{eqnarray}}
\newcommand{\eea}{\end{eqnarray}}
\newcommand{\dxx}{\partial_{X\hbox{\hskip-.1cm}X}}
\newcommand{\xx}{_{X\hbox{\hskip-.1cm}X}}

\begin{document}

\title{
Nonlinear dynamics in one dimension:\\
On a criterion for coarsening  and its temporal law }

\author{Paolo Politi}
\email{Paolo.Politi@isc.cnr.it}
\affiliation{Istituto dei Sistemi Complessi,
Consiglio Nazionale delle Ricerche, Via Madonna del Piano 10,
50019 Sesto Fiorentino, Italy}

\author{Chaouqi Misbah}
\email{chaouqi.misbah@ujf-grenoble.fr} \affiliation{Laboratoire de
Spectrom\'etrie Physique, CNRS, Univ. J. Fourier, Grenoble 1, BP87,
F-38402 Saint Martin d'H\`eres, France}

\date{\today}

\begin{abstract}
We develop a general criterion about coarsening for  a  class of
nonlinear evolution equations describing one dimensional
pattern-forming systems. This criterion  allows one to discriminate
between the situation where a coarsening process takes place and the
one where the wavelength is fixed in the course of time. An
intermediate scenario may  occur, namely  `interrupted coarsening'.
The power of the criterion on which a brief account has been given
[P. Politi and C. Misbah, Phys. Rev. Lett. {\bf 92 }, 090601
(2004)], and which we extend here to more general equations, lies in
the fact that the statement about the occurrence of coarsening, or
selection of a length scale, can be made by only inspecting the
behavior of the branch of steady state periodic solutions. The
criterion states that coarsening occurs if $\lambda'(A)>0$ while a
length scale selection prevails if $\lambda'(A)<0$, where $\lambda$
is the wavelength of the pattern and $A$ is the amplitude of the
profile (prime refers to differentiation). This criterion is established
thanks to the analysis of the phase diffusion equation of the
pattern. We connect the phase diffusion coefficient $D(\lambda)$
(which carries a
kinetic information) to $\lambda'(A)$, which refers to a pure
steady state property. The relationship between kinetics and the
behavior of the branch of steady state solutions is established
fully analytically for several classes of equations. Another important
and new result which emerges here is that the exploitation of the phase
diffusion coefficient enables us to determine in a rather
straightforward manner the dynamical coarsening exponent. Our
calculation, based on the idea that $|D(\lambda)|\sim\lambda^2/t$, 
is exemplified on several nonlinear equations, showing
that the exact exponent is captured. We are not aware of another
method that so systematically provides the coarsening exponent. Contrary
to many situations where the one dimensional character has proven
essential for the derivation of the coarsening exponent, this idea
can be used, in principle, at any dimension. Some speculations about
the extension of the present results are outlined.
\end{abstract}



\maketitle

\section{Introduction}
\begin{figure}
\includegraphics[width=7cm,clip]{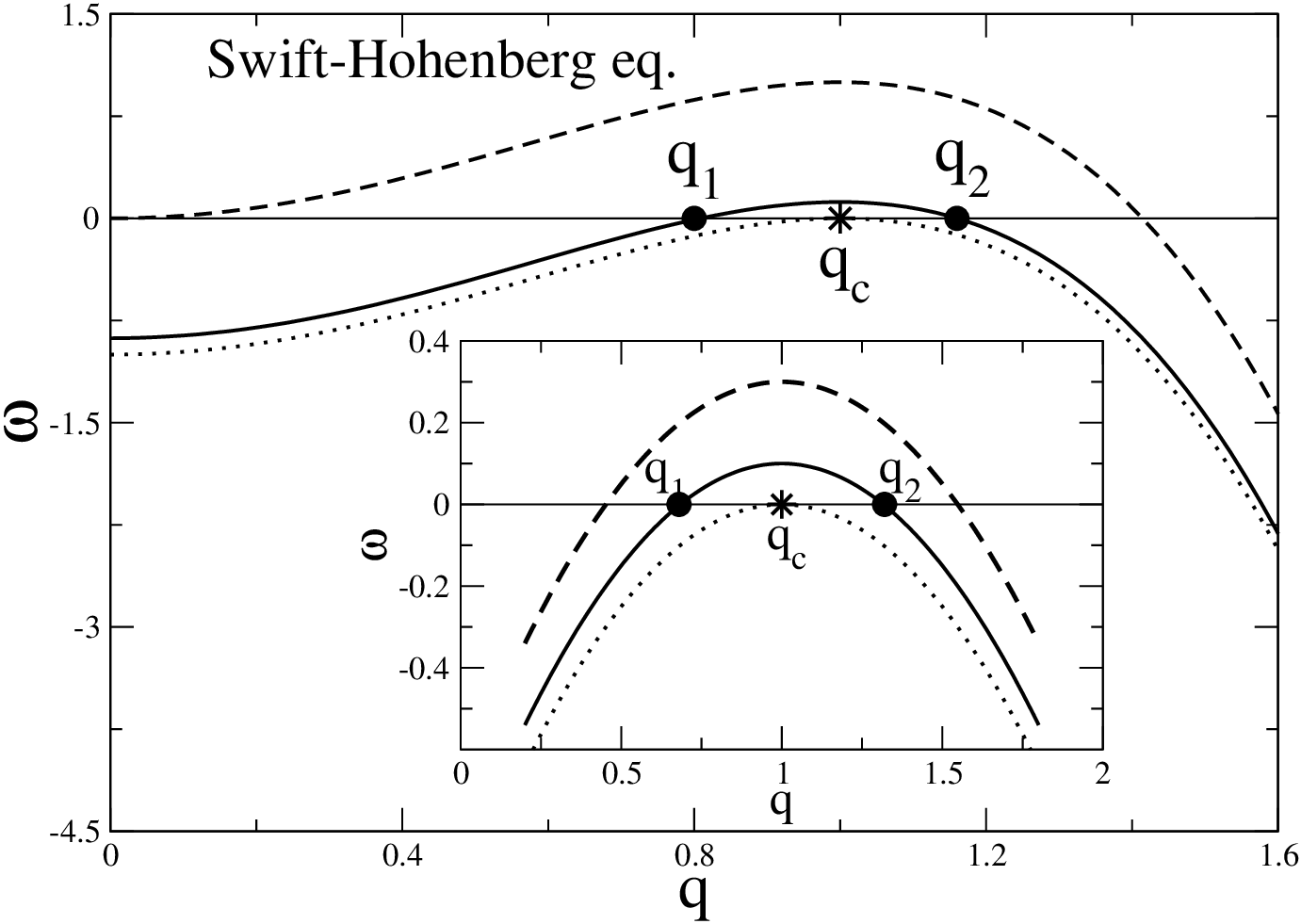}
\caption{Inset: Dispersion curve (\protect\ref{dispI}):
$\omega =\delta - (q-q_c)^2$.
There is loss of stability at a critical wavenumber $q_c$,
when $\delta=0$ (dotted line). For $\delta>0$, the unstable band extends
from $q=q_1$ to $q=q_2$ (full line). With increasing $\delta$ the
unstable region widens (dashed line).
The parabolic shape of $\omega(q)$ is an approximation, valid close
to its maximum. This applies, e.g., to the dispersion curve
(\protect\ref{dispSH}) of the Swift-Hohenberg eq. (see the main figure):
$\omega = \delta - (1-q^2)^2$ [$q_c=1$]. When $\delta=1$ (dashed line)
the unstable band extends down to $q=0$ and $\omega(q)$ resembles
the dispersion curve of the Cahn-Hilliard eq. (see (\protect\ref{dispII})
and Fig.~\protect\ref{fig_w_CH-GL}).
}
\label{fig_w_SH}
\end{figure}

Pattern formation is ubiquitous in nature, and especially for
systems which are brought away from  equilibrium. Examples  are
encountered in hydrodynamics, reaction-diffusion systems,
interfacial problems, and so on. There is now an abundant literature
on this topic~\cite{cross93,patterns_book}.
Generically, the first stage of
pattern formation is the loss of stability of the homogeneous
solution against a spatially periodic modulation. This generally
occurs at a critical value of a control parameter, $\mu=\mu_c$
(where $\mu$ stands for the control parameter) and at a critical
wavenumber $q=q_c$. The dispersion relation about the homogeneous
solution (where perturbations are sought as $e^{iqx+\omega t}$), in
the vicinity of the critical point assumes, in most of
pattern-forming systems, the following parabolic form
(Fig.~\ref{fig_w_SH}, inset)
\begin{equation}
\omega=\delta -  (q-q_c)^2
\label{dispI}
\end{equation}
where  $\delta$ is proportional to $(\mu-\mu_c)$. For $\delta<0$, $\omega<0$
for all $q's$ and the homogeneous state is stable. Conversely, for
$\delta>0$ there is a band of wavevectors 
$\Delta q \equiv (q-q_c) =\pm
\sqrt{\delta}$ corresponding to unstable modes (Fig. 1), so that
infinitesimal perturbations grow exponentially with time until
nonlinear effects can no longer be ignored. In the vicinity of the
bifurcation point ($\delta=0$) only the principal harmonic with
$q=q_c$ is unstable, while all other harmonics are stable. For
example, Rayleigh-B\'enard convection, Turing systems, and so on,
fall within this category, and their nonlinear evolution equation is
universal in the vicinity of the bifurcation point. If the field of
interest (say a chemical concentration) is written as $A(x,t)
e^{iq_c x}$, where $A$ is a complex slowly varying amplitude,  then
$A$ obeys the canonical equation
\begin{equation}
\partial _t A =A+\partial _{xx} A-|A|^2A
\label{LGA}
\end{equation}
where it is supposed that the coefficient of the cubic term is
negative to ensure a nonlinear saturation. Because the band of
active modes is narrow and centered around the principal harmonic,
no coarsening can occur, and the pattern will select a given length,
which is often close to that of the linearly fastest growing mode.
However, the amplitude equation above exhibits a phase instability,
known under the Eckhaus instability~\cite{cross93}, stating that
among the band of allowed states, $|\Delta q|=\sqrt{\delta}$, only
those modes whose wavevectors  satisfy $|\Delta
q|<\sqrt{\delta/3}$ are stable
 with respect to a wavelength modulation.

\begin{figure}
\includegraphics[width=7cm,clip]{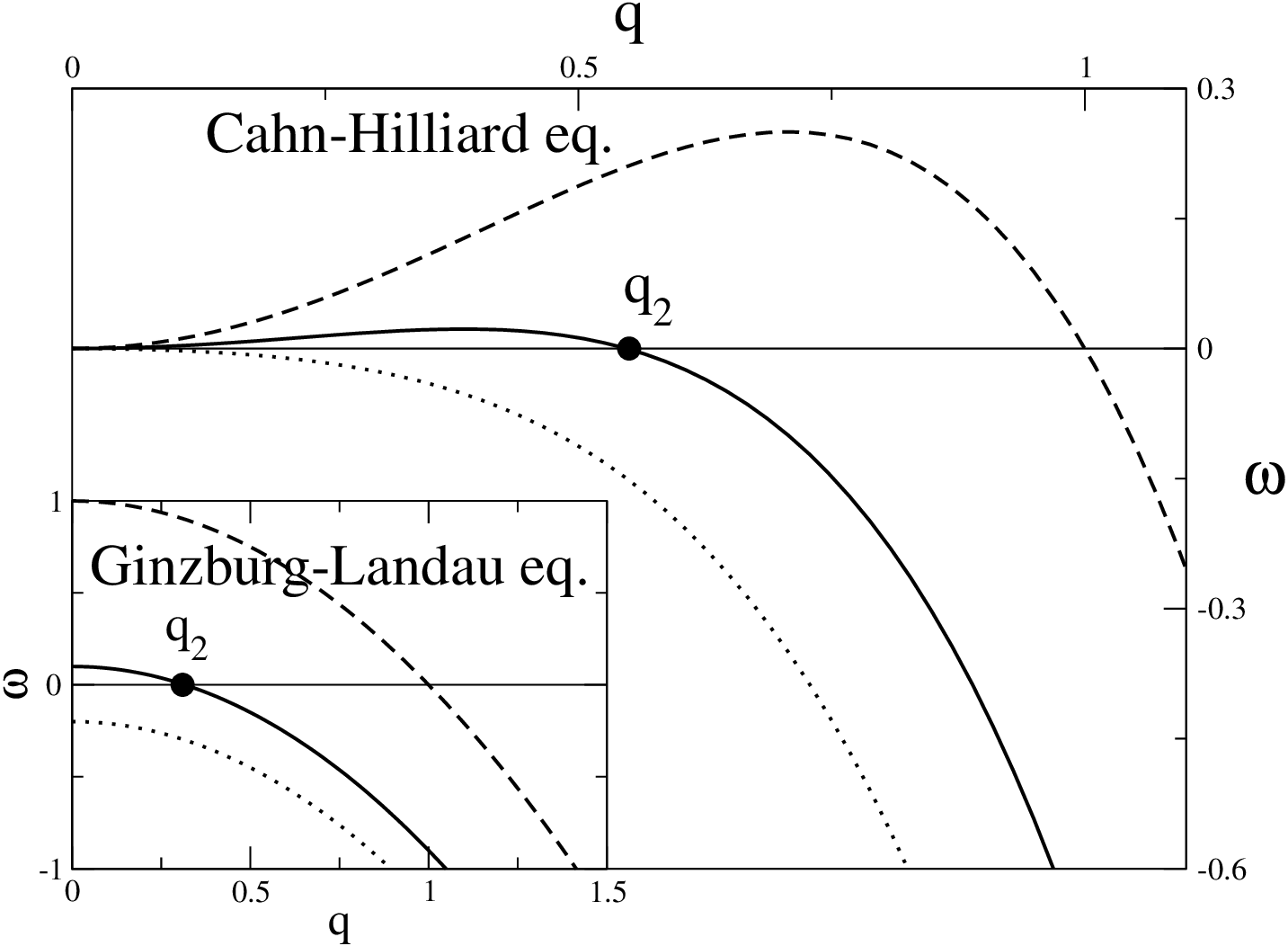}
\caption{Main: dispersion curve (\protect\ref{dispII}), valid, e.g., for
the Cahn-Hilliard equation: $\omega=\delta q^2-q^4$. Dotted line:
below threshold ($\delta<0$). Full line: just above threshold.
Dashed line: well above threshold. The vanishing of $\omega(0)$ for any
$\delta$ is a consequence of the traslational invariance of the
CH eq. in the `growth' direction. Inset: The Ginzburg-Landau eq. is one case
where such invariance is absent and the dispersion curve has the form
(\protect\ref{dispIIp}): $\omega=\delta -q^2$.
}
\label{fig_w_CH-GL}
\end{figure}

 There are many other situations where the bifurcation wavenumber
 $q_c\rightarrow 0$ and therefore a separation of a slow
 amplitude and a fast oscillation is illegitimate. Contray to 
the case (\ref{dispI}),
 where the field can be written as $A(x,t) e^{iq_cx}$ with $A$ being
 supposed to vary slowly in space and time, if $q_c\to 0$ the
 supposed fast oscillation, $e^{iq_cx}$, becomes slow as well and
 a separation of $A$ does not make a sense anymore.
 In this case, a generic form of the dispersion relation is
(Fig.~\ref{fig_w_CH-GL}, main)
\begin{equation}
\omega=\delta q^2 -q^4 .
\label{dispII}
\end{equation}
A third situation is the one where the dispersion relation takes
the form (Fig.~\ref{fig_w_CH-GL}, inset)
\begin{equation}
\omega=\delta -q^2 .
\label{dispIIp}
\end{equation}
 In both cases, Eqs.~(\ref{dispII},\ref{dispIIp}),
the instability occurs for $\delta>0$, and  the
band of unstable modes extends from $q=0$ to $q=\sqrt{\delta}$. That is
to mean, there is an infinite number of unstable harmonics:
if $\bar q$ is the wavenumber of an unstable mode, then also $\bar
q/2, \bar q/3, \dots$ are unstable. Examples that fall in this
category are numerous~\cite{Misbah94}: the air-liquid interface in a
thin film falling on an inclined plane, flame fronts, step dynamics
in step flow growth, sand ripples, and so on, or simply the
Ginzburg-Landau equation (\ref{LGA}), which corresponds to case
(\ref{dispIIp}). 

Dispersion relation (\ref{dispII}) has an extra factor of $q^2$
which is often due to a conservation constraint (see also below). 
Because of the dispersion form, constant plus a quadratic term,
(\ref{dispIIp}) might formally resemble (\ref{dispI}). 
However, an important caution must be taken: in
(\ref{dispI}) it must be remembered that $q$ should remain close to
$q_c$, so that only one harmonic is active, while in
($\ref{dispIIp}$) no such a restriction is made, and therefore $q$
can be as close as possible to zero, leading to a highly nonlinear
dynamics. 

Other types of dispersion relations which may arise, and which are
worth of mention, correspond to situations where $\omega=\delta
|q| -q^2$, or $\omega=\delta |q| -|q|^3$, leading also to a
vigorous mode mixing, for the same reasons evoked above. The
occurrence of a non analytic dispersion relation with $|q|$ is a
consequence of long range interactions~\cite{Kassner02}.
If the unstable band extends down to $q=0$, the appropriate
form of the evolution equation is not an amplitude equation for a
slowly varying function $A$, but rather a partial nonlinear
differential equation, or an integro-differential equation, for the
full field of interest,\footnote{In case (\protect\ref{dispIIp})
an equation similar to (\protect\ref{LGA}) may arise, but it
describes the full field and not just the envelope.} say $h(x,t)$
if one has in mind a front profile.

A prominent example of
a partial differential equation is the
Kuramoto-Sivashinsky~\cite{Nepo,Kuramoto,Sivashinsky} (KS) equation
\begin{equation}
\partial _t u=-\partial _{xx}u -\partial _{xxxx}u+ u\partial _{x}u
\label{KS}
\end{equation}
which leads to spatio-themporal chaos.
Note that by setting $u=\partial_x h$ we obtain an equivalent form of this
equation, namely $\partial _t h=-\partial _{xx}h -\partial _{xxxx}h+
(\partial _{x}h)^2/2$. This equation  arises in several contexts:
liquid films flowing down an inclined plane~\cite{Nepo},
flame fronts~\cite{Sivashinsky},
step flow growth of a crystal surface~\cite{Misbah94}.

Complex dynamics such as chaos, coarsening, etc~$\dots$,
are naturally expected if modes of arbitrarily large wavelength
are unstable. However, these dynamics may occur
for systems characterized by the dispersion relation (\ref{dispI}) as well, 
if the system is further driven away from the
critical point (i.e., if $q_2 \gg q_1$, see Fig.~\ref{fig_w_SH}) because
higher and higher harmonics become active.
We may expect, for
example, coarsening to become possible up to a total wavelength of
the order of $2\pi/q_1$.

For systems which are at global equilibrium the nonlinearity
$u\partial _x u$ is not allowed, and a prototypical equation having
the dispersion relation (\ref{dispII}) is the Cahn-Hilliard equation
\begin{equation}
\partial _t u=-\partial _{xx}[u +\partial _{xx}u -u^3] .
\label{CH}
\end{equation}
The linear terms are identical to the KS one, and the difference
arises from the nonlinear term. Note that if dynamics is not subject
to  a conservation constraint, $(-\partial _{xx})$ on the right hand
side is absent, and the dispersion relation is given by Eq.~(\ref{dispIIp}).
The resulting equation is given by (\ref{LGA})
for a real $A$ and it is called real Ginzburg-Landau (GL) equation
or Allen-Cahn equation.

The KS  equation, or its conserved form (obtained by applying
$\partial_{xx}$ on the right hand side), was suspected for a long
time to arise as the generic nonlinear evolution equation for
nonequilibrium systems (the quadratic term is non variational in
that it can not be written as a functional derivative) whenever a
dispersion relation is of type (\ref{dispII}). Several recent studies,
especially in Molecular Beam Epitaxy (MBE), have revealed an
increasing evidence for the occurrence of completely new types of
equations, with a variety of dynamics: besides chaos, there are
ordered multisoliton~\cite{Misbah96,Sato96} solutions,
coarsening~\cite{Review}, freezing of the wavelength accompanied by a
perpetual increase of the amplitude~\cite{OPL}. Moreover, equations
bearing strong resemblance with each other~\cite{Paulin} exhibit a
completely different dynamics. Thus it is highly desirable to
extract some general criteria that allow one to discriminate between
various dynamics.

A central question that has remained open so far, and which has been
the subject of a recent brief exposition~\cite{Politi04}, was the
understanding of the general conditions under which dynamics should
lead to coarsening, or rather to a selection of a length scale. In
this paper we shall generalize our proof presented
in~\cite{Politi04} to a larger number of classes of nonlinear equations,
for which the same general criterion applies: the sign of the phase
diffusion coefficient $D$ is linked to a property of the steady state
branch. More precisely, the sign of $D$ is shown to be the opposite
of the sign of $\lambda'(A)$, the derivative of the wavelength $\lambda$
of the steady state with respect its amplitude $A$.
Therefore, coarsening occurs if (and only if) the wavelength increases
with the amplitude.

Another important new feature  that constitutes  a subject of this
paper, is the fact that the exploitation of the phase diffusion
coefficient $D(\lambda)$ will allow us to derive analytically the coarsening
exponent, i.e. the law according to which the wavelength of the
pattern increases in time.
For all known nonlinear equations whose dispersion relation has
the form (\ref{dispII}) or (\ref{dispIIp})
and display coarsening,
we have obtained the exact value of the coarsening exponent,
and we predict exponents for other non exploited yet equations. An
important point is that this is expected to work at any dimension.
Indeed, the derivation of the phase equation can be done
in higher dimension as well.
If our criterion, based on the idea that $|D(\lambda)|\sim\lambda^2/t$,
remains valid at higher dimensions, it should
become a precious tool for a straightforward derivation of the
coarsening exponent at any dimension.

\section{The phase equation method}
\label{sec_phase-eq}

\subsection{Generality}
\label{sec_general}

Coarsening of an ordered pattern occurs if steady state periodic
solutions are unstable with respect to wavelength fluctuations. The
phase equation method~\cite{Kevorkian}
allows to study in a perturbative way the
modulations of the phase $\phi$ of the pattern. For a periodic
structure of period $\lambda$, $\phi=qx$, where $q=2\pi/\lambda$ is
a constant. If we perturb this structure, $q$ acquires a space and
time dependence and the phase $\phi$ is seen to satisfy a diffusion
equation, $\partial_t\phi=D\partial_{xx}\phi$. The quantity $D$,
called phase diffusion coefficient, is a function of the steady
state solutions and its sign determines the stable ($D>0$) or
unstable ($D<0$) character of a wavelength perturbation.

A negative value of $D$ induces a coarsening process,\footnote{
In principle a negative $D$ could entail also a decreasing of $\lambda$ (splitting).
However, this is inconsistent with the result $d{\cal F}/d\lambda <0$
(see Sec.~\ref{sec_pot}) and with the stability of the
flat interface at small length scales.}
whose typical
time and length scales are related by $|D(\lambda)| \sim \lambda^2/t$,
as simply derived from the solution of the phase diffusion equation:
this relation allows to find the coarsening law $\lambda(t)$.
Therefore, the phase equation method not only allows to determine
if certain classes of partial differential equations (PDE)
display coarsening or not; it also allows
to find the coarsening laws, when $D<0$.
In the rest of this section, we are going to offer a short exposition
of the phase
equation method without referring to any specific PDE. Explicit evolution
equations will be treated in the next sections, with some calculations
relegated to the appendix.

Let us consider a general PDE of the form\footnote{Coarsening
scenarios are not affected by the presence of noise, which is not
taken into account throughout the article.}
\be
\partial_t u(x,t) = \tilde{\cal N}[u]
\label{eq_nl} \ee where $\tilde{\cal N}$ is an unspecified nonlinear
operator, which is assumed not to depend explicitly on space and
time. $u_0(x)$ is a periodic steady state solution: $\tilde{\cal
N}[u_0]=0$ and $u_0(x+\lambda)=u_0(x)$.

When studying the perturbation of a steady state, it is useful to
separate a fast spatial variable from slow time and space
dependencies. The stationary solution $u_0$ does not depend on time
and it has a fast spatial dependence, which is conveniently
expressed through the phase $\phi=qx$. Once we perturb the
stationary solution, 
\be u = u_0 + \epsilon u_1 + \dots  ,
\label{u_exp} 
\ee 
the wavevector $q=\partial_x\phi$ gets a slow
space and time dependence: $q=q(X,T)$, where $X=\epsilon x$ and
$T=\epsilon^\alpha t$. Because of the diffusive character of the
phase variable, the exponent $\alpha$ is equal to two. Space and
time derivatives now read 
\bse
\label{new_der}
\bea
\partial_x &=& q\partial_\phi + \epsilon\partial_X \\
\partial_t &=& \epsilon (\partial_T\psi)\partial_\phi
\eea
\ese
where the second order term in the latter equation
($\epsilon^2\partial_T$) has been neglected. Finally,
along with the phase $\phi$
it is useful to introduce the slow phase $\psi(X,T)=\epsilon\phi(x,t)$,
so that $q=\partial_X\psi$.

Replacing the $u-$expansion (\ref{u_exp}) and the derivates (\ref{new_der})
with respect to the new variables in Eq.~(\ref{eq_nl}), we find an
$\epsilon -$expansion which must be vanished term by term.
The zero order equation is trivial, $\tilde{\cal N}_0[u_0]=0$: this equation
is just the rephrasing of the time-independent equation in terms of the
phase variable $\phi$ (the subscript in $\tilde{\cal N}_0$ means that
Eqs.~(\ref{new_der}) have been applied at zero order in $\epsilon$,
i.e. $\partial_x = q\partial_\phi$).

The first order equation is more complicated, because both the operator
$\tilde{\cal N}$ and the solution $u$ are $\epsilon -$expanded.
On very general grounds, we can rewrite $\partial_t u(x,t) = \tilde{\cal N}[u]$
as
\be
\epsilon (\partial_T\psi)\partial_\phi u_0 =
(\tilde{\cal N}_0 + \epsilon\tilde{\cal N}_1)[u_0+\epsilon u_1]
\ee
where $\tilde{\cal N}_1$ comes from first order contributions to
the derivatives (\ref{new_der}). If we use the Fr\'echet
derivative~\cite{Zwillinger},
$\tilde{\cal L}_0$, defined through the relation
\be
\tilde{\cal N}_0[u_0+\epsilon u_1]=
\tilde{\cal N}_0[u_0] + \epsilon \tilde{\cal L}_0[u_1] + O(\epsilon^2)
\ee
we get
\be
\tilde{\cal L}_0[u_1] = (\partial_T\psi)\partial_\phi u_0 -
\tilde{\cal N}_1[u_0] \equiv g(u_0,q,\psi) .
\label{eq_ord1}
\ee

At first order, therefore, we get an heterogeneous linear equation
(the Fr\'echet derivative of a nonlinear operator is linear).
The translational invariance of the operator $\tilde{\cal N}$ guarantees
that $\partial_\phi u_0$ is solution of the homogeneous equation:
according to the Fredholm alternative theorem~\cite{Fredholm},
a solution for the
heterogeneous equation may exist only if $g$ is orthogonal to the
null space of the adjoint operator $\tilde{\cal L}^\dagger$.
In simple words, if $\tilde{\cal L}^\dagger [v]=0$, $v$ and $g$
must be orthogonal. This condition, see Eq.~(\ref{eq_ord1}), reads
\be
\langle v,\partial_\phi u_0\rangle \partial_T\psi =
\langle v, \tilde{\cal N}_1[u_0]\rangle ,
\ee
where\footnote{Sometimes we may also write $\langle f(\phi)\rangle$
to mean $(2\pi)^{-1}\oint d\phi f$.}
$\langle f,g\rangle = (2\pi)^{-1}\int_0^{2\pi} d\phi f^*g$.

It happens that $\tilde{\cal N}_1[u_0]$ is proportional to
$\partial_X q =\dxx\psi$, and the previous equation
has the form of a diffusion equation for the phase $\psi$,
\be
\partial_t \psi = D \dxx\psi .
\ee

\subsection{Applications}
\label{sec_app}

\subsubsection{The generalized Ginzburg-Landau equation}
\label{sec_GL}

The (real) Ginzburg-Landau equation is written as
\be
\partial_t u = (u-u^3) + u_{xx}
\label{eq_sGL}
\ee
whose linear spectrum, for an excitation $u(x,t)=
\exp(\omega t +iqx)$ is $\omega(q)=1-q^2$. This equation is the
prototype for the evolution of a nonconserved order parameter
with two equivalent stable solutions, $u=\pm 1$. Starting from
the trivial solution $u=0$, we have a linear instability leading
to a logarithmically slow coarsening process~\cite{Langer}.

This equation can be easily generalized to
\be
\partial_t u = B(u) + G(u)\, u_{xx} \; ,
\label{eq_GL}
\ee
which will therefore be called {\it generalized Ginzburg-Landau} (gGL) equation.
If $B(u)\simeq u$ and $G(u)\simeq 1$ for small $u$, the linear spectrum
is unmodified, but the nonlinear behavior can be totally different,
depending on the full (positive) expressions of $B(u)$ and $G(u)$. Steady states
are determined by the relation $u_0''(x) = -B(u_0)/G(u_0)$, so
they correspond to the trajectories of a classical particle moving
under the force $-B(u_0)/G(u_0)$.

Now, let us apply the expansions (\ref{u_exp}) for the order parameter
and (\ref{new_der}) for the derivatives to Eq.~(\ref{eq_GL}). The first
and second spatial derivatives can also be written as
\bse
\bea
\partial_x &=& q\partial_\phi + \epsilon\psi\xx\partial_q \\
\partial_{xx} &=& q^2\partial_{\phi\phi} + \epsilon\psi\xx
(2q\partial_q +1)\partial_\phi
\eea
\ese

As anticipated in the previous section, the zero and first order
equations read $\tilde{\cal N}_0[u_0] = 0$ and $\tilde{\cal L}_0[u_1]
= g$, where
\be
\tilde{\cal N}_0[u_0]=B(u_0)+G(u_0)q^2\partial_{\phi\phi}u_0
\ee
is the nonlinear operator defining the gGL equation,
\be
\tilde{\cal L}_0[u_1]=[B'(u_0)+G'(u_0)q^2(\partial_{\phi\phi}u_0)
+G(u_0)q^2\partial_{\phi\phi}]u_1
\ee
is its Fr\'echet derivative, and
\be
g=(\partial_T\psi)\partial_\phi u_0 - (\dxx\psi)
G(u_0)(2q\partial_q +1)\partial_\phi u_0 \; .
\ee

Because of translational invariance, $\tilde{\cal L}_0[\partial_\phi u_0]
=0$. Its adjoint is easily found to be
\be
\tilde{\cal L}^{\dagger}_0[v]=q^2\partial_{\phi\phi}[vG(u_0)]
+[B'(u_0)+G'(u_0)q^2(\partial_{\phi\phi}u_0)]v .
\ee
If we define $w=vG(u_0)$, the equation $\tilde{\cal L}^{\dagger}_0[v]=0$
is identical to $\tilde{\cal L}_0[w]=0$, so that we can choose 
$w=\partial_\phi u_0$ and $v=\partial_\phi u_0/G(u_0)$.

The orthogonality condition between $v$ and $g$ reads
\be
(\partial_T\psi)\langle v,\partial_\phi u_0\rangle -
(\dxx\psi)\langle v,G(u_0)(2q\partial_q +1)\partial_\phi u_0\rangle =0
\ee
and replacing the explicit expression for $v$, we get the phase
diffusion equation
\be
\partial_T\psi = D \dxx\psi
\ee
with
\be
D = \frac{\partial_q\langle q(\partial_\phi u_0)^2\rangle}{%
\left\langle {(\partial_\phi u_0)^2\over G(u_0)}\right\rangle}
\equiv {D_1\over D_2} .
\label{eq_D-GL}
\ee

Assuming a positive $G$, the sign of $D$ is fixed by the increasing
or decreasing character of $\langle q(\partial_\phi u_0)^2\rangle$
with the wavevector $q$. Reversing to the old variable $x$,
\be
\langle q(\partial_\phi u_0)^2\rangle =
{1\over 2\pi}\int_0^\lambda dx (u_0')^2 = {J\over 2\pi}
\ee
where $J$ is the well known action variable, whose derivative with respect
to the `energy' of the particle gives the period $\lambda$.
The following relations are easily established:
\be
D_1 = {1\over 2\pi}{\partial J\over\partial q} = -{\lambda^3\over
4\pi^2}\left({\partial\lambda\over\partial E}\right)^{-1} =
-{\lambda^3\over 4\pi^2}{B(A)\over G(A)}
\left({\partial\lambda\over\partial A}\right)^{-1}
\label{eq_D1-GL}
\ee
where $A$ is the amplitude of the oscillation, i.e. the (positive)
maximal value for $u_0(x)$.

If $G(u)\equiv 1$, a compact formula for $D$ is
\be
D = - {\lambda^2 B(A)\over J (\partial_A\lambda)} ~~~~ G(A)\equiv 1 .
\label{eq_D-GL_semp}
\ee

In conclusion, a coarsening process occurs ($D<0$) if the wavelength
of the steady states increases with increasing their amplitude.
In App.~\ref{app_lambda} we make some general remarks on the
behavior of $\lambda(A)$ for several different potentials.

\subsubsection{The generalized Cahn-Hilliard equation}
\label{sec_CH}

The Cahn-Hilliard equation is the conserved `version' of the
Ginzburg-Landau equation,

\be
\partial_t u = -\partial_{xx} [ (u-u^3) + u_{xx} ] .
\label{eq_sCH}
\ee

The spatial average of the order parameter is time independent,
$d\langle u\rangle/dt=0$, and the linear spectrum is $\omega(q)=
q^2-q^4$: it therefore has a maximum at a finite value
$q_u=1/\sqrt{2}$, called the most unstable wavevector.
The linear regime corresponds to an exponential unstable growth of
such mode, with a rate $\omega(q_u)$, followed by a logarithmic
coarsening.\footnote{The coarsening
of the nonconserved (Ginzburg-Landau) and conserved
(Cahn-Hilliard) models differ if noise is present:
$\lambda(t)\sim t^{1/2}$ in the former case and
$\lambda(t)\sim t^{1/3}$ in the latter case.}

The above equation can be made of wider application by considering
the following {\it generalized Cahn-Hilliard} (gCH) equation
\be
\partial_t u = - C(u)\partial_{xx} [ B(u) + G(u) u_{xx} ] .
\label{eq_CH}
\ee

In Sec.~\ref{sec_n-cg}
we will discuss thoroughly the coarsening of this
class of models, because of its relevance for the crystal growth of
vicinal surfaces.\footnote{A vicinal 
surface is a surface which is slightly miscut from
a high-symmetry orientation. It looks like a flight of stairs
with steps of atomic height separating large terraces.}
In that case, the local height $z(x,t)$ of the
steps satisfies the equation
\be
\partial_t z = -\partial_x\{ B(m)+G(m)\partial_x [C(m)\partial_x m]\}
\label{eq_cg}
\ee
where $m=\partial_x z$. If we pass to the new variable
$u(m)=\int_0^m ds C(s)$ and take the spatial derivative of the above
equation, we get the gCH equation (\ref{eq_CH}).
It is worthnoting that steady states are given by the equation
$B(u_0) + G(u_0) u_0''=j_0$, where $j_0$ is a constant determined by the
condition $\langle u_0\rangle = m_0$ that imposes the (conserved) average
value of the order parameter. If steps are oriented along a high-symmetry
orientation, $m_0=0=j_0$. In the following we are considering this case only,
so the equation determining steady states, $B(u_0) + G(u_0) u_0''=0$,
is the same as for the gGL equation.

If we proceed along the lines explained in Sec.\ref{sec_general} and
keep in mind notations used in Sec.\ref{sec_GL}, the first order equation
in the small parameter $\epsilon$ reads
\bea
 -q^2\partial_{\phi\phi}\tilde{\cal L}_0[u_1]  &&
= (\partial_T\psi){\partial_\phi u_0\over C(u_0)} \\
&& + (\dxx\psi) q^2\partial{\phi\phi}[
G(u_0)(2q\partial_q +1)\partial_\phi u_0] .
\nonumber
\eea

According to the Fredholm alternative theorem, the right hand side must be
orthogonal to the solution $v$ of the equation
\be
(\partial_{\phi\phi}\tilde{\cal L}_0)^\dagger [v] =
\tilde{\cal L}_0^\dagger \partial_{\phi\phi}v =0 .
\ee

According to the results of the previous section, we know that
$\partial_{\phi\phi}v =(\partial_\phi u_0)/G(u_0)$. The orthogonality
condition now reads
\bea
&&\left\langle v, {\partial_\phi u_0\over C(u_0)}\right\rangle (\partial_T\psi) \\
&+& q^2 \langle v, \partial{\phi\phi}[
G(u_0)(2q\partial_q +1)\partial_\phi u_0]\rangle (\dxx\psi) =0 .
\nonumber
\eea

The quantity multiplying $(\dxx\psi)$ can be rewritten as
\bea
\langle v, \partial{\phi\phi}[G(u_0)(2q\partial_q +1)\partial_\phi u_0
\rangle &=& \nonumber\\
\langle {\partial_\phi u_0\over G(u_o)}, G(u_0) (2q\partial_q +1)
\partial_\phi u_0\rangle &=& \nonumber\\
\partial_q \langle q(\partial_\phi u_0)^2 \rangle &&
\eea
so we finally have $(\partial_T\psi) = D (\dxx\psi)$ with
\be
D =- {q^2\partial_q \langle q(\partial_\phi u_0)^2 \rangle \over
\left\langle v, {\partial_\phi u_0\over C(u_0)}\right\rangle } \equiv
{q^2 D_1 \over \tilde D_2} .
\label{eq_D-CH}
\ee

In App.~\ref{app_sign-D} we prove that the denominator  $\tilde D_2$ is always
positive. If $C$ and $G$ are (positive) constants the proof is straightforward,
because $-\langle v\partial_\phi u_0\rangle=\langle(\partial_\phi v) u_0\rangle
= \langle u_0^2\rangle$. The diffusion coefficient (\ref{eq_D-CH})
for the gCH equation is therefore similar to the diffusion coefficient
(\ref{eq_D-GL}) for the nonconserved gGL equation: their sign is determined
by the increasing or decreasing character of $\lambda(A)$, the wavelength
of the steady state, with respect to its amplitude. The $q^2$ term
in the numerator of (\ref{eq_D-CH}) is evidence of the conservation law,
i.e., of the second derivative $\partial_{xx}$ in Eq.~(\ref{eq_CH}).
The denominators $D_2$ and $\tilde D_2$ differ: this is irrelevant for the
sign of $D$, but it is relevant for the coarsening law.

If $C(u)\equiv G(u)\equiv 1$, formulas simplify:
$D_1=-\lambda^3 B(A)/ 4\pi^2 (\partial_A\lambda)$ and
$\tilde D_2=\langle u_0^2\rangle=I/\lambda$, where
$I=\oint u_0^2(x)$ has the same role as $J=\oint (\partial_x u_0)^2$
in the nonconserved model. Putting everything together we obtain
\be
D = - {\lambda^2 B(A)\over I (\partial_A\lambda)} ~~~~
C(A)\equiv G(A)\equiv 1 .
\label{eq_D-CH_semp}
\ee

\subsubsection{The generalized Swift-Hohenberg equation}
\label{sec_SH}

Both the (generalized) GL and CH equations have a linear spectrum
whose unstable band extends to $q=0$, so that active modes of
arbitrarily large wavelength exist. In the Swift-Hohenberg equation
we can tune a parameter $\delta$ so that to change the unstable band.
The standard form of the equation is
\bea
\partial_t u &=& \delta u -(1+\partial_{xx})^2 u -u^3 \label{eq_SH} \\
&=& -\partial_x^4 u -2\partial_x^2 u - (1-\delta)u -u^3
\nonumber
\eea

Linear stability analysis for a single harmonic $u(x,t)=
\exp(\omega t +iqx)$ gives the spectrum
\be
\omega(q) = \delta - (1-q^2)^2
\label{dispSH}
\ee
and for positive $\delta$ there is a finite unstable band ($\omega(q)>0$)
extending from $q_1=\sqrt{1-\sqrt{\delta}}$ to $q_2=\sqrt{1+\sqrt{\delta}}$.
For $\delta=1$, the unstable band is the interval $(0,\sqrt{2})$.
The most unstable wavevector is $q_u=1$ for any $\delta$.
For small $\delta$ the unstable band is narrow; in fact, for $\delta<0.36$,
$q_1 > q_2/2$ and period doubling is not allowed. In other
words studying coarsening for the Swift-Hohenberg equation close to the
threshold $\delta=0$ is not very interesting: nonetheless we will
write the phase diffusion equation for {\it any} $\delta$ and for a
generalized form of the Swift-Hohenberg equation as well.

The zero order equation is easy to write
\be
\tilde{\cal N}_0[u_0] \equiv -q^4\partial^4_\phi u_0 -2 q^2\partial^2_\phi u_0
- (1-\delta) u_0 - u_0^3 = 0
\ee
and the first order equation has the expected form $\tilde{\cal L}_0[u_1]=g$,
where
\be
\tilde{\cal L}_0 \equiv -q^4\partial^4_\phi -2 q^2\partial^2_\phi
- (1-\delta)  - 3u_0^2
\ee
is the Fr\'echet derivative of $\tilde{\cal N}_0$ and
\bea
&&g\equiv (\partial_T\psi) \partial_\phi u_0 \\
&&+ 2(\dxx\psi)
[(2q^3\partial_q +3q^2)\partial^3_\phi u_0 +(2q\partial_q +1)\partial_\phi u_o] .
\nonumber
\eea

The operator $\tilde{\cal L}_0$ is self-adjoint, so the solution
of the homogeneous equation $\tilde{\cal L}_0^\dagger[v]=0$ is
immediately found, because of the translational invariance of
$\tilde{\cal N}_0$ along $x$: $v=\partial_\phi u_0$. We therefore have
\bea
(\partial_T\psi)\langle(\partial_\phi u_0)^2\rangle =\\
- (\dxx\psi)[
\langle\partial_\phi u_0 (4q^3\partial_q+6q^2)\partial^3_\phi u_0\rangle
\nonumber\\
+2\langle\partial_\phi u_0(2q\partial_q+1)\partial_\phi u_0\rangle] .
\nonumber
\eea

It is easy to check that both terms appearing in square brackets on the
right hand side can be written as $\partial_q(\dots)$:
\bse
\bea
\langle\partial_\phi u_0 (4q^3\partial_q+6q^2)\partial^3_\phi u_0\rangle &=&
-\partial_q\langle 2q^3(\partial^2_\phi u_0)^2\rangle \\
\langle\partial_\phi u_0(2q\partial_q+1)\partial_\phi u_0\rangle] &=&
\partial_q \langle q(\partial_\phi u_0)^2\rangle
\eea
\ese
so that the phase diffusion coefficients reads
\be
D = {\partial_q[2q^3\langle(\partial^2_\phi u_0)^2\rangle -2q\langle
(\partial_\phi u_0)^2\rangle ] \over
\langle (\partial_\phi u_0)^2\rangle } .
\label{eq_D-SH}
\ee

Now let us generalize this result to the equation
\be
\partial_t u = \sum_{n>0} c_n \partial_x^n u + P(u)
\ee
where $n$ is assumed even and $P(u)$ an odd function, in order
to preserve the symmetries $x\to -x$ and $u\to -u$. 
Therefore $n=2k$, with
$k\ge 1$. We report here only the final result for the phase diffusion
coefficient:
\be
D = {1\over 2\langle (\partial_\phi u_0)^2\rangle} \sum_n (-1)^{k+1}
nc_n\partial_q \langle q^{n-1}(\partial_\phi^{n/2}u_0)^2\rangle
\label{eq_D-gSH}
\ee

The standard Swift-Hohenberg equation corresponds to $c_2=-2$ and
$c_4=1$. The quantity $(-1)^{k+1}nc_nq^{n-1}/2$ therefore gives
$2q^3$ for $n=4$ ans $-2q$ for $n=2$, as shown by Eq.~(\ref{eq_D-SH}).

\section{The coarsening exponent}
\label{sec_n}

We now want to use the results obtained in the previous section for
the phase diffusion coefficient $D$ in order to get the coarsening law
$\lambda(t)$. In one dimensional systems, noise may be relevant and change
the coarsening law. In the following we will restrict our analysis
to the deterministic equations.

A negative $D$ implies an unstable behavior of the phase
diffusion equation, $\partial_t\psi=-|D|\partial_{xx}\psi$, which displays an
exponential growth (we have reversed to the old coordinates for the sake
of clarity): $\psi=\exp(t/\tau)\exp(2\pi ix/\lambda)$, with 
$(2\pi)^2|D|=\lambda^2/\tau$
(in the following the time scale $\tau$ will just be written as $t$).
The relation $|D(\lambda)|\approx\lambda^2/t$ will therefore be used
to obtain the coarsening law $\lambda(t)$: it will be done for
several models displaying the scenario of perpetual coarsening
(i.e., $\lambda'(A)>0$ for diverging $\lambda$).

\subsection{The standard Ginzburg-Landau and Cahn-Hilliard models}
\label{sec_n-GL-CH}

It is well known~\cite{Langer}
that in the absence of noise, both the nonconserved
GL equation (\ref{eq_sGL}) and the conserved CH equation
(\ref{eq_sCH}) display logarithmic coarsening, $\lambda(t)\approx\ln
t$. Let us remind that steady states correspond to the trajectories
of a classical particle moving in the potential $V(u)=u^2/2 -u^4/4$.
The wavelength of the steady state, i.e. the oscillation period,
diverges as the amplitude $A$ goes to one. This limit corresponds
to the `late stage' regime in the dynamical problem, and the
profile of the order parameter is a sequence of kinks and antikinks.
The kink (antikink) is the stationary solution $u_+(x)$ ($u_-(x)$)
which connects $u=-1$ ($u=1$) at $x=-\infty$ to $u=1$ ($u=-1$) at $x=\infty$,
$u_\pm(x)=\pm\tanh(x/\sqrt{2})$.
As coarsening proceeds, kinks and antikinks annihilate.

It is convenient to introduce the new variable
$Q(x)=1-u(x)$. In the mechanical analogy, if the particle leaves
(with zero velocity) from $u(0)=A=1-Q_0$, the quantity $Q(x)=1-u(x)$
starts to grow exponentially (this follows from the
$\tanh(x/\sqrt{2})$ solution,\footnote{This may also be directly seen
by expanding the differential equation $\partial _{xx} u+u-u^3=0$
about $u=1$. Expansion to leading order in $Q=1-u$
yields $\partial _{xx} Q-2Q=0$, from which the solution $Q=Q_0
e^{\sqrt{2}x}$ follows.} $Q(x)\simeq Q_0\exp(\sqrt{2}x)$). The
particle passes a diverging time close to $x=\pm 1$ because
$x=\pm 1$ correspond to the potential
maxima where the velocity of the particle vanishes while its
position remain finite. Consequently,
it is straightforward to write\footnote{It
is possible to calculate more rigorously the period of oscillation
from the exact relation $\lambda =2\int_{-A}^A
du/(\partial _x u)$, which can be expressed in terms of an elliptic
integral. In the limit $A\rightarrow 1$ it diverges logarithmically
as $-\ln (1-A)$.}
that $\lambda\approx\ln(1/Q_0)\approx
-\ln(1-A)$ and calculate the derivative $\partial_A\lambda\approx
(1-A)^{-1} \approx \exp\lambda$.

In order to apply Eq.~(\ref{eq_D-GL_semp}) and find $D$ for the
nonconserved model, we also
need $B(A)=A-A^3\approx2\exp(-\lambda)$ and $J$; the latter quantity
is the classical action and it is easy to understand that it keeps
finite in the limit $A\to 1$, because it is the area in the phase space
enclosed in the limiting trajectory. So, the relation
$|D(\lambda)|\approx \lambda^2/t$ gives
$\partial_A\lambda/B(A) \approx t$
and finally
\be
\lambda(t) \approx \ln t   ~~\hbox{~~~~standard GL eq} .
\label{eq_n-GL}
\ee

If we compare Eq.~(\ref{eq_D-GL_semp}) with Eq.~(\ref{eq_D-CH_semp}),
we find that the phase diffusion coefficient $D$ for the nonconserved
model is equal to that for the conserved model, once $J=\oint
(\partial_x u_0)^2$ has been replaced by $I=\oint (u_0)^2$, which is
of order $\lambda$ because $u_0$ is almost everywhere equal to one.
This replacement is irrelevant in our case, because the relation
$\exp(2\lambda)\sim t$, giving rise to Eq.~(\ref{eq_n-GL})
in the nonconserved case, is replaced by $\lambda\exp(2\lambda)\sim t$
in the present, conserved case, so that
a logarithmic coarsening is found again:
\be
\lambda(t) \approx \ln t   ~~\hbox{~~~~standard CH eq} .
\label{eq_n-CH}
\ee

\subsection{Other models with uniform stable solutions}
\label{sec_n-uss}

The only features of the standard GL and CH models which determine
the coarsening laws (\ref{eq_n-GL}) and (\ref{eq_n-CH}) are the
presence of symmetric maxima in $V(u)$ for finite $u=\bar u=1$ and
the quadratic behavior close to the maxima, $V(u)\approx V(\bar u) -
a (u-\bar u)^2$. For any generalization of such equations,
$\partial_t u = B(u) +u_{xx}$ and $\partial_t u =
-\partial_{xx}[B(u) +u_{xx}]$, if $V(u)=\int du B(u)$ still has such
properties, the coarsening will be logarithmic independently of the
details of $B(u)$. Since a quadratic behavior close to a maximum is
general, we can conclude that a logarithmic coarsening is common to
most of nonconserved and conserved models where $B(u)$ vanishes for
finite $u$.

In this section we investigate situations where the assumption
that $V(u)$ has a quadratic behavior close to the maxima is relaxed.
They corresponds to very special situations, nonetheless they are
interesting in principle because $\lambda(t)$ is no more logarithmic
and because these models constitute  a bridge between the standard
GL/CH models and the models without uniform stable solutions,
discussed in the next section. Let us now consider a class of models
where $V(u)$ has a minimum in $u=0$ (linearly unstable profile) and
two symmetric maxima in $u=\pm 1$, with $V(u)\simeq V(1)
-a|u-1|^\beta$ ($\beta>2$) close to $u=1$. Again, we use
$Q(x)=1-u(x)$ and $Q_0=1-A$, $A$ being the amplitude of the
oscillation.

The wavelength is given by \be \lambda \sim \int_{Q_0}^1
{dQ\over\sqrt{Q^\beta-Q_0^\beta}} \sim{1\over Q_0^{{\beta\over
2}-1}} \ee and it is easily found that
$J\sim 1$, $I\sim\lambda$, $\partial_A\lambda\sim
Q_0^{-\beta/2}$, $B(A)\sim Q_0^{\beta -1}$.

\subsubsection{The nonconserved case}

The relation \be {\lambda^2\over t} \sim |D| = {\lambda^2 B(A)\over
J(\partial_A\lambda)} \label{gimondi} \ee gives \be t\sim
{Q_0^{-\beta/2}\over Q_0^{\beta -1}} \sim
\lambda^{3\beta-2\over\beta-2} \ee so that $\lambda(t) \sim t^n$
with $n=(\beta-2)/(3\beta-2)$.

\subsubsection{The conserved case}

If the order parameter is conserved, we simply need to replace $J\sim 1$
with $I\sim\lambda$ in Eq.~(\ref{gimondi}), so as to obtain
\be
t \sim \lambda \lambda^{3\beta-2\over\beta-2}
\sim \lambda^{4\beta-4\over\beta-2} .
\ee

The coarsening exponent is therefore equal to $n=(\beta-2)/(4\beta-4)$.
We observe that in the limit $\beta\to 2$ we recover logarithmic
coarsening ($n=0$) both in the nonconserved and conserved case, as it should
be. We also remark that in the opposite limit $\beta\to\infty$ we get
$n=1/3$ in the nonconserved case and $n=1/4$ in the conserved
case, which make a bridge towards the models discussed in the next section.

\subsection{Models without uniform stable solutions}
\label{sec_n-alpha}

The models considered in the previous subsection have
uniform stable solutions, $u=\pm 1$: the linear instability of
the trivial solution $u=0$ leads to the formation of domains
where the order parameter is alternatively equal to $\pm 1$,
separated by domain walls, called kinks and antikinks.
This property is related to the fact that $B(u)=u-u^3$
vanishes for finite $u$ (up to a sign, $B(u)$ is the force in the
mechanical analogy for the steady states).

In the following we are considering a modified class of models,
where $B(u)$ vanishes in the limit $u\to\infty$ only, so that
the potential $V(u)=\int du B(u)$ does not have maxima
at finite $u$.
Therefore, it is not possible to define `domains'
wherein the order parameter takes a constant value.
These models~\cite{Torcini}, which may be relevant for the
epitaxial growth of a high-symmetry crystal surface~\cite{Review},
are defined as follows ($\alpha >1$):
\bea
&&\partial_t u = B(u) + u_{xx} \\
&&\partial_t u = -\partial_{xx} [B(u) + u_{xx}] \label{eq_alpha-con} \\
&& B(u) = {u\over(1+u^2)^\alpha} \\
&& V(u) = -{1\over 2(\alpha -1)} {1\over  (1+u^2)^{\alpha -1}}
\eea

Steady states correspond to periodic oscillations of a particle in
the potential $V(u)$ around the minimum in $u=0$. With increasing
the amplitude $A$, the energy of the particle goes to zero as $E\sim
-A^{-(2\alpha-2)}$ and the motion can be split into two parts: the
motion in a finite region around $u=0$, which does not depend on
$A$, and the motion for large $u$, where $V(u)\sim 1/u^{(2\alpha
-2)}$ and simple dimensional analysis can be used. As an example,
let us evaluate the wavelength through the relation (where the
`acceleration' is proportional to  the `force' in the mechanical
language) 
\be 
{A\over\lambda^2} \sim {1\over A^{2\alpha-1}}
~~\Longrightarrow ~~ \lambda\sim A^\alpha . 
\ee

In an analogous way, we can evaluate the action
$J\sim\lambda(A/\lambda)^2\sim\lambda^{{2\over\alpha}-1}$,
which is slighty
more complicated, because the asymptotic contribution vanishes
for $\alpha >2$: in this case the finite, constant contribution coming
from the motion within the `close region' dominates. Therefore,
$J\sim \lambda^{{2\over\alpha}-1}$
if $\alpha\le 2$ and $J\sim 1$ if $\alpha > 2$.
As for the quantity $I$, appearing in the expression (\ref{eq_D-CH_semp})
for $D$, conserved case, we simply have \
$I\sim \lambda A^2\sim \lambda^{{2\over\alpha}+1}$. Finally,
$B(A)\sim\lambda^{{1\over\alpha}-2}$.

\subsubsection{The nonconserved case}

  From Eq.~(\ref{eq_D-GL_semp}) we find
\bse
\bea
D \sim {\lambda^2 \lambda^{{1\over\alpha}-2}\over\lambda^{{2\over\alpha}-1}
\lambda^{1-{1\over\alpha}}} \sim 1 ~~~~~\alpha <2 \\
D \sim {\lambda^2 \lambda^{{1\over\alpha}-2}\over
\lambda^{1-{1\over\alpha}}} \sim
\lambda^{{2\over\alpha}-1} ~~~~~\alpha >2
\eea
\ese

The phase diffusion coefficient is constant for $\alpha$ smaller than two,
which means that $\lambda(t)\sim t^{1\over 2}$. For $\alpha$ larger than
two, $\lambda^{{2\over\alpha}-1}\sim \lambda^2/t$ gives
$\lambda(t)\sim t^{\alpha\over 3\alpha -2}$. We can sum up our results,
$\lambda(t) \sim t^n$, with
\bse
\bea
n = \hbox{${1\over 2}$} ~~~~\alpha < 2 \\
n = {\alpha\over 3\alpha -2} ~~~~\alpha > 2
\eea
\ese

The coarsening exponent varies with continuity from $n=1/2$
for $\alpha<2$ to $n=1/3$ for $\alpha\to\infty$.
These results confirm what had already been found by one of us with a
different approach~\cite{Torcini}.

\subsubsection{The conserved case}

  From Eq.~(\ref{eq_D-CH_semp}) we have
\be
D \sim {\lambda^2 \lambda^{{1\over\alpha}-2}\over\lambda^{{2\over\alpha}+1}
\lambda^{1-{1\over\alpha}}} \sim {1\over\lambda^2}
\ee
and therefore
\be
\lambda(t) \sim t^{1\over 4} .
\label{eq_n-con}
\ee

The constant coarsening exponent $n=1/4$ clashes with numerical
results found in Ref.~\cite{Torcini},
$n=1/4$ for $\alpha <2$ and $n=\alpha/(5\alpha-2)$ for
$\alpha>2$.
The opinion of the authors of Ref.~\cite{Torcini}
is that for $\alpha>2$ a crossover should exist from
$n=\alpha/(5\alpha-2)$ to $n=1/4$, the correct asymptotic
exponent. Details and supporting
arguments will be given elsewhere~\cite{erratum}.

\subsection{Conserved models for crystal growth}
\label{sec_n-cg}

It is interesting to consider a model of physical interest which
belongs to the class of the {\it full generalized} Cahn-Hilliard equations,
meaning that all the functions $B(u)$, $C(u)$ and $G(u)$
appearing in (\ref{eq_CH}) are not trivial.
The starting point is Eq.~(\ref{eq_cg}),
$$
\partial_t z = -\partial_x\{ B(m)+G(m)\partial_x [C(m)\partial_x m]\}
$$
which describes the meandering
of a step, or---more generally---the meandering of a train of steps
moving in phase. $z$ is the local displacement of the step
with respect to its straight position and $m=\partial_x z$
is the local slope of the step.

We do not give details here about the origin of the previous
equation, which is presented in~\cite{Paulin}, but just write the
explicit form of the functions $B,G$ and $C$: 
\bse
\label{eq_BGC}
\bea 
B(m) = {m\over 1+m^2}
, ~~
G(m) = {1+\beta\sqrt{1+m^2}\over (1+\beta)(1+m^2)} ,
\label{eq_BG} \\
C(m) = {1+c(1+m^2)(1+2m^2)\over (1+c)(1+m^2)^{3/2} }
\label{eq_C}
\eea
\ese
and define the meaning of the two adimensional, positive parameters
appearing there: $\beta$ is the relative strength between the two
relaxing mechanisms, line diffusion and terrace diffusion;
$c$ is a measure of the elastic coupling between steps.

If we pass
to the new variable $u=\int_0^m ds\;C(s)$, we get Eq.~(\ref{eq_CH}),
\be
\partial_t u = - C(u)\partial_{xx} [ B(u) + G(u) u_{xx} ]
\tag{\protect\ref{eq_CH}}
\ee
whose steady states, for high-symmetry steps, are given by the equation
$u_{xx} = -B(u)/G(u)$. In App.~\ref{app_cg} we study
the potential $V(u)=\int du [B(u)/G(u)]$ and the dynamical
scenarios emerging from $\partial_A\lambda$. We give here the results only.

\begin{figure}
\includegraphics[width=7cm,clip]{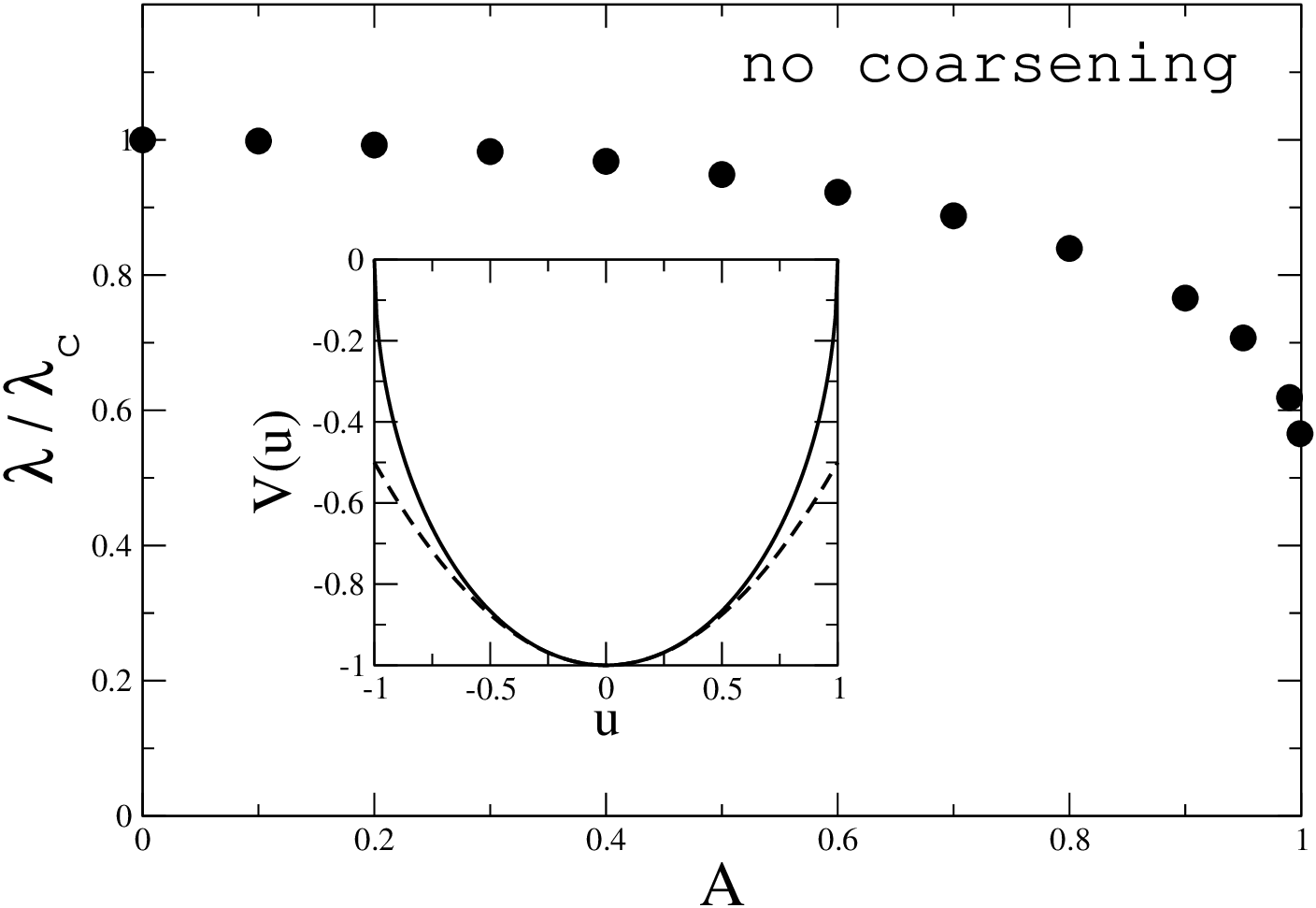}
\caption{$\lambda(A)$ for Eq.~(\protect\ref{eq_CH}), $c=0$ and $\beta=0$
(see Eq.~(\protect\ref{eq_BGC})). $V(u)=-\sqrt{1-u^2}$
(inset) and $\lambda'(A)<0$ (main, full circles). 
The dashed line is the harmonic potential
$u^2/2$.  If $c=0$, there is no coarsening whatever is $\beta$. }
\label{fig_0_0}
\end{figure}

\begin{figure}
\includegraphics[width=7cm,clip]{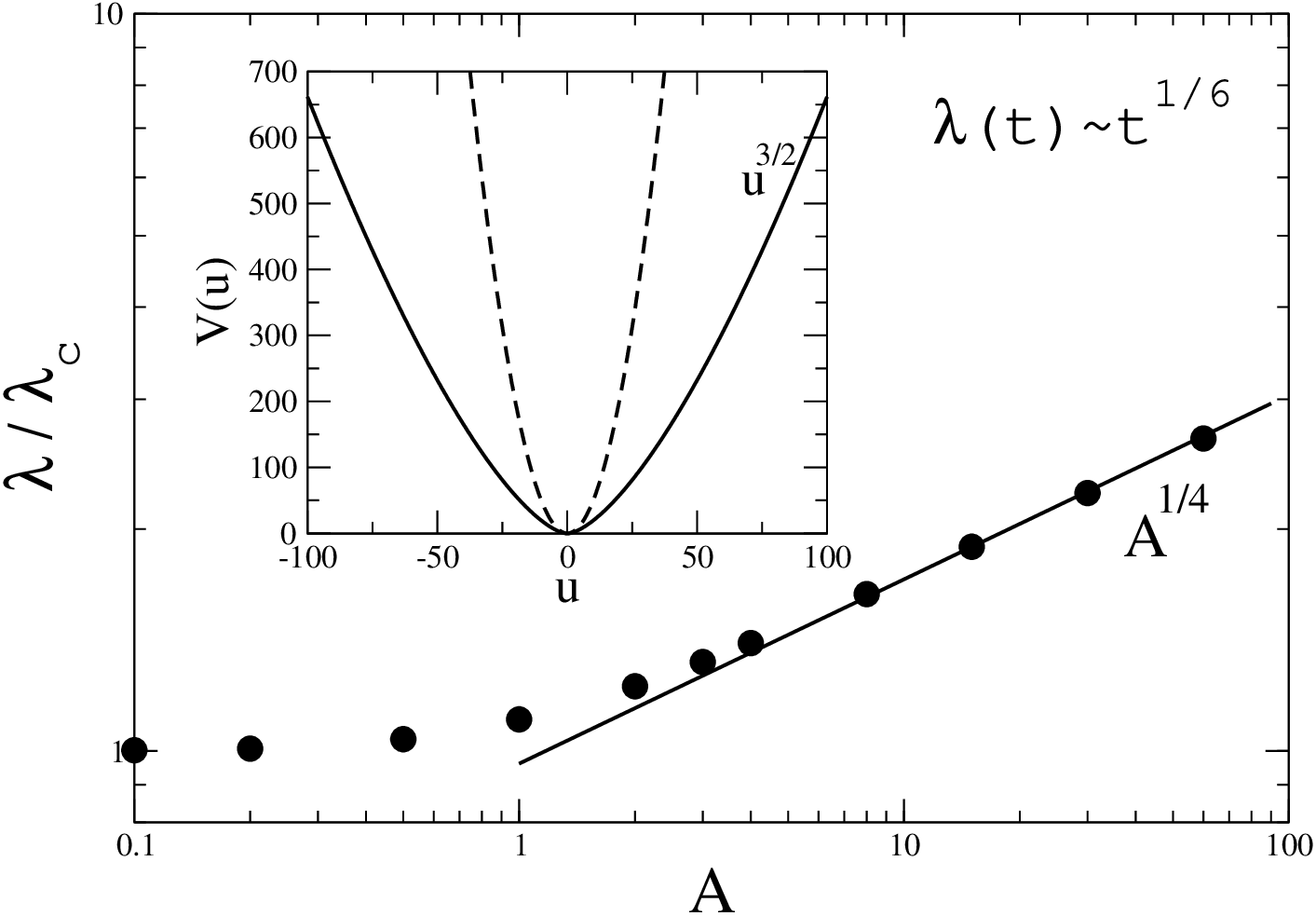}
\caption{$\lambda(A)$ for Eq.~(\protect\ref{eq_CH}), $c=\infty$ and $\beta=0$.
Inset: $V(u)=\sqrt{1+m^2}(2m^2-1)/3$, with $m^2=(\sqrt{1+4u^2}-1)/2$.
$V(u)\approx |u|^{3/2}$ for large $u$ and $\lambda'(A)>0$ (main, full
circles). For large amplitude, $\lambda\approx A^{1/4}$ (main, full line).
The dashed line in the inset is the harmonic potential, $u^2/2$.
If $c>0$ there is coarsening, whose law depends on $\beta$.
For $\beta=0$, $\lambda(t)\approx t^{1/6}$. }
\label{fig_inf_0}
\end{figure}

\begin{figure}
\includegraphics[width=7cm,clip]{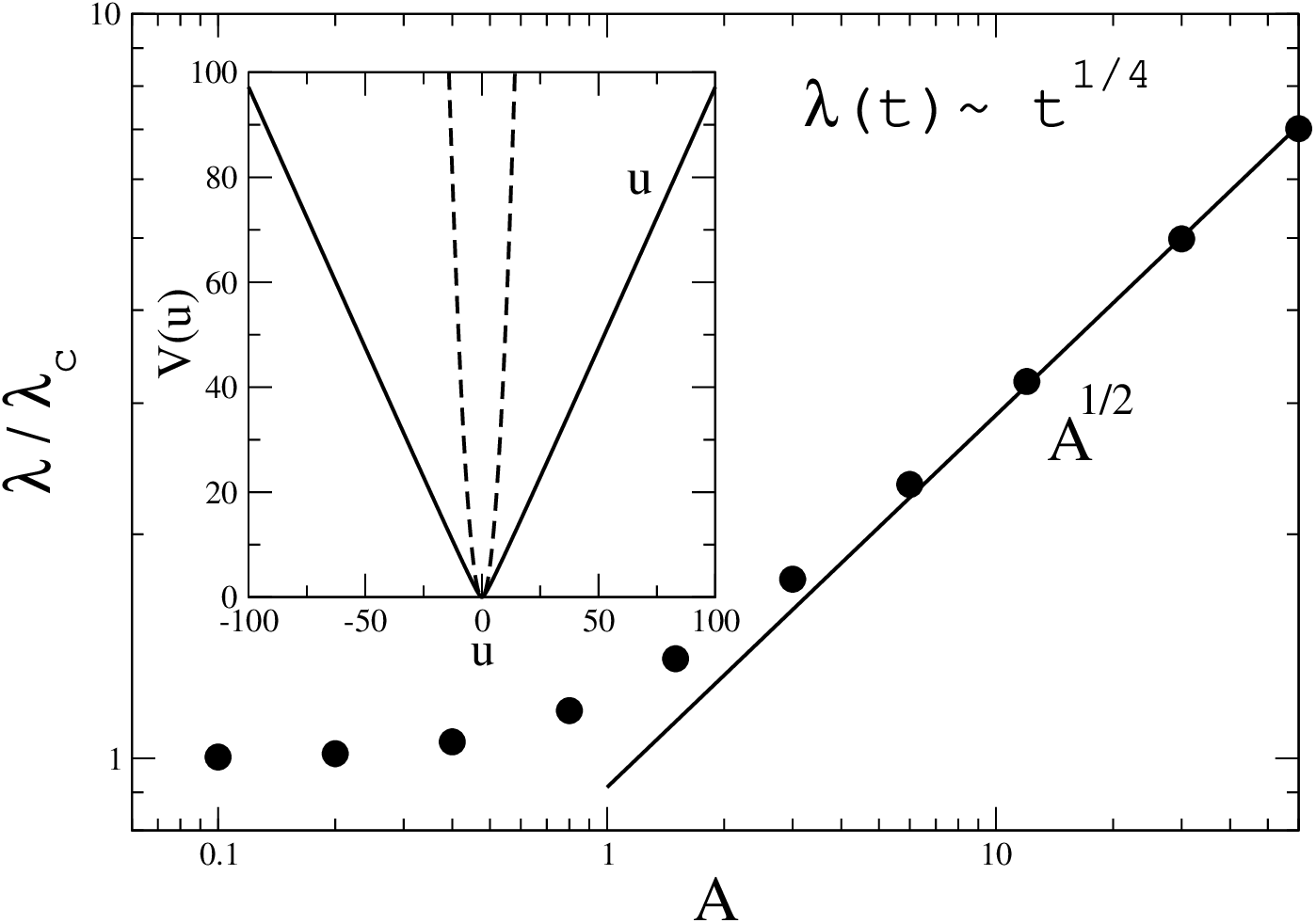}
\caption{$\lambda(A)$ for Eq.~(\protect\ref{eq_CH}), $c=\infty$ and $\beta=\infty$.
Inset: $V(u)=m^2-(1/2)\ln(1+m^2)$, with $m^2=(\sqrt{1+4u^2}-1)/2$.
$V(u)\approx |u|$ for large $u$ and $\lambda'(A)>0$ (main, full
circles). For large amplitude, $\lambda\approx \sqrt{A}$ (main, full line).
The dashed line in the inset is the harmonic potential, $u^2/2$.
For $\beta>0$, $\lambda(t)\approx t^{1/4}$. }
\label{fig_inf_inf}
\end{figure}

If $c=0$ (see Fig.~\ref{fig_0_0}), $\partial_A\lambda < 0$,
while there is asymptotic coarsening if $c>0$
(see Figs.~\ref{fig_inf_0},\ref{fig_inf_inf}). Asymptotic coarsening means
that $\partial_A\lambda > 0$ for large enough $A$: according to the values of
$c$ and $\beta$, $\lambda(A)$ may be always increasing or it may
have a minimum followed by $\partial_A\lambda > 0$: the distinction
between the two cases is not relevant for the dynamics and it will
not be further considered. Let us now determine the asymptotic
behavior of all the relevant quantities, when $c>0$.

In the limit of large $m$, we have $C(m)\sim m$ and $B(m)\sim 1/m$.
As for $G$, $G(m)\sim 1/m^2$ and $G(m)\sim 1/m$, for $\beta=0$
and $\beta\ne 0$ respectively. Since $u\sim\int^m ds\; C(s)\sim m^2$,
$m\sim\sqrt{u}$ so that $C(u)\sim\sqrt{u}$, $B(u)\sim 1/\sqrt{u}$ and
$G(u)\sim 1/u$ ($\beta=0$) or $G(u)\sim 1/\sqrt{u}$
($\beta\ne 0$). The potential $V(u)$ has the form $V(u)\sim u^{3/2}$
($\beta=0$, Fig.~\ref{fig_inf_0}) or $V(u)\sim u$ ($\beta\ne 0$,
Fig.~\ref{fig_inf_inf}). The wavelength (see App.~\ref{app_lambda}) is
$\lambda\sim A^{1/4}$ for $\beta=0$ (Fig.~\ref{fig_inf_0}) and
$\lambda\sim\sqrt{A}$ for $\beta\ne 0$ (Fig.~\ref{fig_inf_inf}).
Similar and straightforward relations can be determined and the following
general expression for the phase diffusion coefficient is established,
\be
D \sim {\sqrt{A}G(A)\over\lambda^2} ~\sim ~ {\lambda^2\over t}
\ee
and the coarsening exponent is finally found to be
\bse
\bea
\lambda(t) \sim t^{1\over 6} & & \beta=0 \label{eq_n-cg1} \\
\lambda(t) \sim t^{1\over 4} & & \beta\ne 0 \label{eq_n-cg2} 
\eea
\ese

These results agree with both the numerical solution and the
heuristic arguments presented in~\cite{Paulin}.

\subsection{Discussion}

In this section we have applied the results of Sec.~\ref{sec_phase-eq}
to find the coarsening law $\lambda(t)$ for some classes of models
displaying asymptotic coarsening, that is to say having a negative phase
diffusion coefficient $D$ for `large' amplitude (if the amplitude $A$
has an upper limit $A_{\hbox{\footnotesize max}}$,
`large' means $A\to A_{\hbox{\footnotesize max}}$).
In particular, in \ref{sec_n-GL-CH}-\ref{sec_n-alpha}
we have considered models entering into the classes
\bse
\bea
\partial_t u = B(u) + u_{xx} &\Rightarrow &
t \sim {J (\partial_A\lambda)\over B(A)} \\
\partial_t u = -\partial_{xx} [B(u) + u_{xx}] &\Rightarrow &
t \sim {I (\partial_A\lambda)\over B(A)}
\eea
\ese
where we have also indicated on the right the relations leading to the
coarsening laws and deriving from $|D|\sim\lambda^2/t$, with $D$ given
in Eqs.~(\ref{eq_D-GL_semp}) and (\ref{eq_D-CH_semp}).

Passing from the standard GL/CH models (Sec.~\ref{sec_n-GL-CH}),
to models where $V(u)$ have (non quadratic) maxima at finite $u$
(Sec.~\ref{sec_n-uss}) and to models where $V(u)$
has no maxima at all at finite $u$ (Sec.~\ref{sec_n-alpha}),
the coarsening exponents change with continuity from $n=0$
(logarithmic coarsening) to $n=1/2$
for the nonconserved models and from $n=0$ to $n=1/4$ for the
conserved models.

The conservation law, as expected, always slows down the coarsening
process.
Formally, this corresponds to replace the action $J$ with the
quantity $I$ in the denominator of $D$. In most cases, $J$ is a
constant while $I$ increases as $\lambda^\kappa$, with $\kappa\ge
1$: a smaller $D$ implies a lower coarsening. We remark that only in
a very special case (models without uniform stable solutions and
$\alpha <2$), $I/J\sim\lambda^2$: when this happens, the double
derivative $\partial_{xx}$---which characterizes the conserved
models---is equivalent (as for the coarsening law) to the factor
$1/\lambda^2$. We stress again that this is an exception,
it is not the rule.

Sec.~\ref{sec_n-cg} has been devoted to a class of conserved models
which are relevant for the physical problem of a growing crystal. In
that case the full expression (\ref{eq_D-CH}) for $D$ must be
considered (the result is reported in
Eqs.~(\ref{eq_n-cg1}-\ref{eq_n-cg2})). It is remarkable that for all
the models we have considered, we found $n\le 1/4$ and $n\le
1/2$ for conserved and nonconserved models, respectively. It
would be interesting to understand how general these inequalities
are.\footnote{The condition $n\le 1/2$ for the nonconserved
models is equivalent to say that $|D(\lambda)|$ does not diverge
with increasing $\lambda$, which seems to be 
a fairly reasonable condition.}

\section{Swift-Hohenberg equation and coarsening}
\label{sec_coar_SH}

Let us start from the standard Swift-Hohenberg equation (\ref{eq_SH}),
\be
\partial_t u = -\partial_x^4 u -2\partial_x^2 u - (1-\delta)u -u^3
\label{again_SH}
\ee
whose linear dispersion curve is $\omega(q)=\delta - (q^2-1)^2$.
The phase diffusion coefficient (see Eq.~(\ref{eq_D-SH})) is
\be
D = 2q{ \partial_q ( \langle u_{xx}^2 \rangle - \langle u_x^2\rangle )\over
\langle u_x^2\rangle }
\ee
and it should be compared to the limiting expression, valid for
$\delta\to 0$,
\be
D_E = -{1\over 2} \left[\omega'' + {(\omega')^2\over\omega}\right]
= -{1\over 2\omega}{d\over dq}(\omega\omega')
\ee
where the subscript $E$ refers to the Eckhaus instability and prime
denotes derivative with respect to $q$.

Steady states satisfy the equation ${\cal L}_{SH}[u] -u^3 =0$, where
${\cal L}_{SH}[u]$ means the linear part of the operator appearing on
the right hand side of Eq.~(\ref{again_SH}). In the one-harmonic
approximation, $u(x)=A(q)\cos(qx)$ and the steady condition writes
$\omega(q)u=u^3$ implying \be \omega(q) = \langle u^2\rangle =
1/2 A^2(q) .
\label{Aqomega}
 \ee

In the same spirit we get $\langle u_x^2\rangle=q^2A^2/2=
q^2\omega$ and $\langle u_{xx}^2\rangle=q^4A^2/2=
q^4\omega$, so that $D$ reads
\be
D={2\over q\omega}{d\over dq}[\omega(q^4-q^2)] .
\ee

Close to the threshold, $\delta\to 0$, $q=1+K$ (with $K\ll 1$),
$\omega(q)=\delta -4K^2$ and $\omega'=-8K$. Finally, we get \be D =
{4\over\omega} {d\over dq} (K\omega) = D_E \label{SH_problem} \ee so
that the expression for $D$ is equal to the well known expression for
$D_E$. It must be noted, see Eq.~(\ref{Aqomega}), that $A$
vanishes at $q=q_2$ and at $q=q_1$, as $\omega$ does, so that
$A$ undergoes a fold singularity at the center of the band. This is
imposed by symmetry, since in the vicinity of threshold the band of
active modes is symmetric. In this case we are in the situation with
the dispersion relation (\ref{dispI}). The phase diffusion
coefficient must also be symmetric with respect to the center, and
therefore it can change sign at the fold, due to this symmetry. Thus
$D$ does not have the sign of $A'$ in this case. Nonetheless, in the
vicinity of $q_2$ the sign of $D$ is still given by $A'$, as shown here
below. Our speculation is that the existence of a fold is likely to
destroy the simple link between  $D$ and $A'$.

\begin{figure}
\includegraphics[width=7cm,clip]{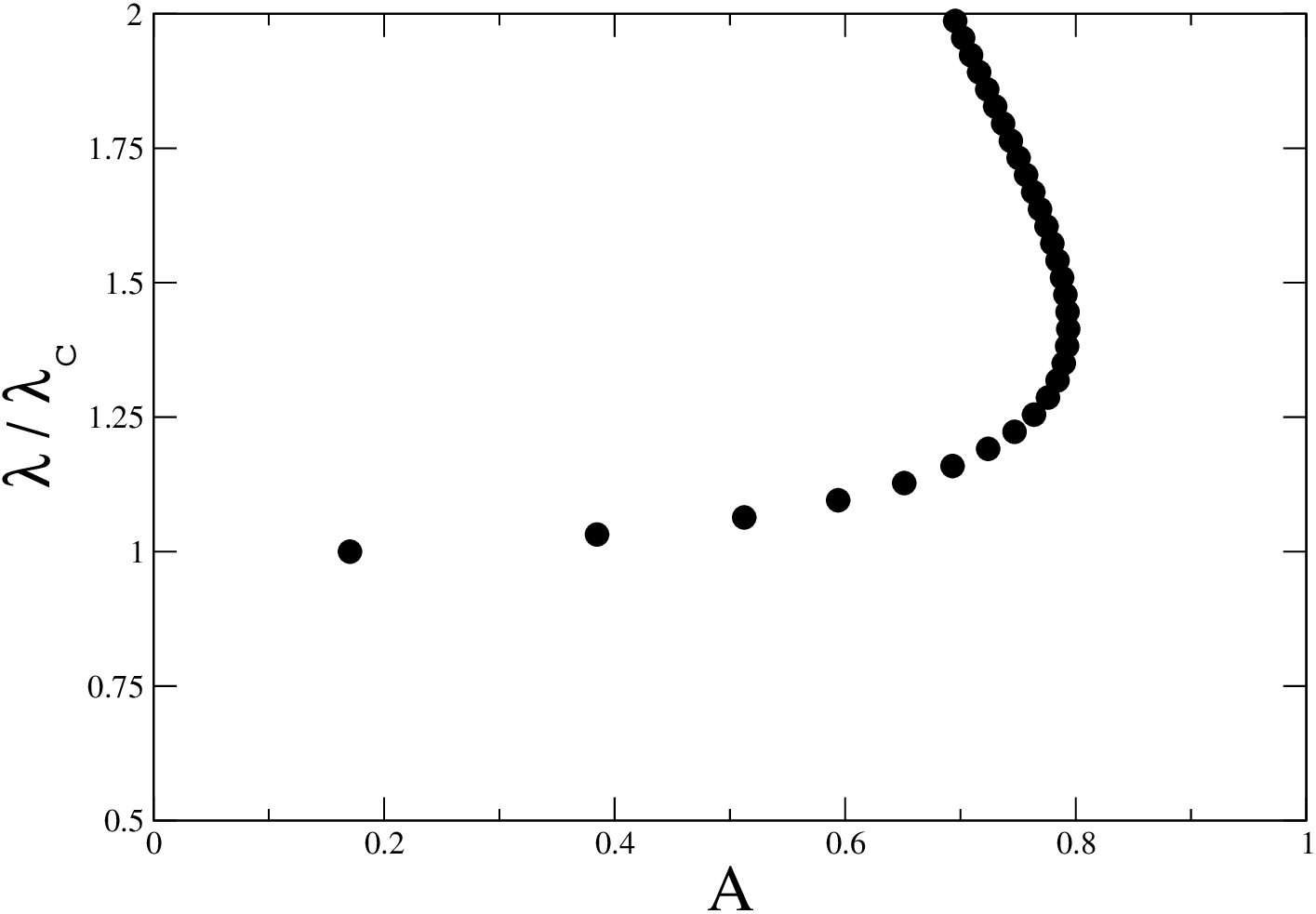}
\caption{$\lambda(A)$ for the Swift-Hohenberg eq.~(\protect\ref{again_SH}), with $\delta=1$.
The amplitude $A$ is defined as $A=[\oint dx\;u^2(x)/\lambda]^{1/2}$.}
\label{fig_SH}
\end{figure}

A meaningful expression for $D$ can be found also for finite
$\delta$, close to $q_2=\sqrt{1+\sqrt{\delta}}$ (let us remind that
$\omega(q_{1,2})=0$ and $q_2 > q_1$). In this limit we get
$\omega'(q)\simeq\omega'(q_2)=4q_2-4q_2^3$ and \be D =
2(q_2^3-q_2){1\over\omega}{d\omega\over dq} = D_E \ee so that $D$ is
equal to a positive quantity times $d\omega/dq$. Since
$\omega(q)=A^2(q)/2$, the sign of $D$ is equal to the sign of
$dA/dq$.  This result follows from a perturbative scheme where use
has been made of the fact that $u\sim \cos(qx)$. This is legitimate
as long as one considers small deviations from the threshold. If
$\delta$ is not small, or if $\delta=1$ we fall in the
dispersion relation (\ref{dispII}). As one deviates from $q_2$
towards the center of the band, higher and higher harmonics become
active, and one should in general find numerically the steady state
solutions in order to ascertain whether $D$ is positive
or negative. In the general case, we have not been able to establish
a link between $D$ and the slope of the steady state branch as done
in the previous sections. Our belief, on which some evidences will
be reported on in the future, is that depending on the class of
equations, it is not always the slope of the steady state solution
that provides direct information on the nonlinear dynamics, but
somewhat a bit more abstract quantities, as we have found, for
example, by investigating the KS equation, another question on which we
hope to report in the near future~\cite{long}. Numerical solutions of the
SH equation in the limit $\delta=1$ reveal a fold singularity in the
branch $\lambda(A)$, as shown in Fig.~\ref{fig_SH}.  

\section{Equations with a potential}
\label{sec_pot}

Some of the equations discussed in Sec.~\ref{sec_app} are derivable
from a potential: it is therefore possible to define a function
${\cal F}$ which is minimized by the dynamics. This is
always the case for the {\it generalized} Ginzburg-Landau equation
(\ref{eq_GL}), which can be written as
\bea
&&\partial_t u = B(u) + G(u)u_{xx} = - G(u){\delta {\cal F}\over\delta u}\\
&& {\cal F}[u] = \int dx \left[
\hbox{${1\over 2}$} (u_x)^2 - V(u) \right]
\eea
where $V(u)$ is the potential entering in the study of the
stationary solutions, i.e. $V'(u)=B(u)/G(u)$. If we evaluate the time
derivative of ${\cal F}$ we find
\be
{d{\cal F}\over dt} = \int dx {\delta {\cal F}\over\delta u} u_t =
-\int dx G(u)\left( {\delta {\cal F}\over\delta u}\right)^2 \le 0 ,
\ee
if $G(u)>0$.

The {\it generalized} Cahn-Hilliard equation (\ref{eq_CH}) can
always be written as
\be
\partial_t u = -C(u)\partial_{xx}[B(u)+G(u)u_{xx}] =
C(u)\partial_{xx}[G(u) {\delta {\cal F}\over\delta u}] .
\ee

If $C(u)= G(u)$, we find
\be
{d{\cal F}\over dt} = -\int dx \left[ \partial_x \left(
C(u) {\delta {\cal F}\over\delta u}\right)\right]^2 \le 0 .
\ee

We now want to evaluate ${\cal F}$ for the steady states.
The pseudo free-energy ${\cal F}$ is nothing but the integral
of the Lagrangian function ${\cal L}$ for the mechanical analogy
defining the stationary solutions. If $E=(u_x)^2/2 +V(u)$
is the `energy' in the mechanical analogy and $J$ is the action,
\be
{\cal F}[u_\lambda(x)] = \ell\left( {J\over\lambda} - E\right)
\ee
where $\ell$ is the length of the one dimensional ($1d$) interface.
We have also made explicit the dependence on $\lambda$ of the
stationary solutions, $u_\lambda(x)$. We want to determine
the dependence on $\lambda$ of the pseudo free-energy,
\be
{d{\cal F}\over d\lambda} =\ell {d\over d\lambda}
\left( {J\over\lambda} - E\right) =
-\ell {J\over\lambda^2} < 0
\ee
where we have used that $\partial_E J=\lambda$.
This result shows that stationary solutions with a greater wavelength
{\it always} have a lower energy. Since dynamics minimizes ${\cal F}$,
this result supports the idea that coarsening occurs when $\partial_A
\lambda >0$.

If the curve $\lambda(A)$ has extrema, there are different steady
states with the same value of $\lambda$: let us consider two of such
states, separated by only one extremum, $\lambda(A_1)=\lambda(A_2)
=\lambda_0$. We ask what state has the lower free energy.
If ${\cal F}_{1,2}=
{\cal F}[u_{\lambda(A_{1,2})}]$ and we label the different steady states
with $E=V(A)$ rather than with $A$, we get
\bea
{ {\cal F}_2 - {\cal F}_1 \over \ell }
&=& {J(E_2)-J(E_1)\over\lambda_0} - (E_2-E_1)\nonumber\\
&=& {1\over\lambda_0}\int_{E_1}^{E_2} dE\lambda(E) - (E_2-E_1)\nonumber \\
&=& {E_2-E_1\over\lambda_0} (\bar\lambda -\lambda_0)
\eea
where $\bar\lambda$ is the average $\lambda$ in the interval $(E_1,E_2)$.

Therefore, if $\lambda(A)$ has a maximum between $A_1$ and $A_2$,
${\cal F}_1<{\cal F}_2$; if $\lambda(A)$ has a minimum
between $A_1$ and $A_2$, ${\cal F}_2<{\cal F}_1$. So, if two steady states
have the same wavelength, the state with lower free-energy is always the
state corresponding to a positive value of $\partial_A\lambda$.

Let us resume what we found in this section. For the gGL equations and
a class of the gCH equations ($C(u)=G(u)$), there is a functional
${\cal F}$ which is minimized by the dynamics, $d{\cal F}/dt\le 0$;
if $u_\lambda(x)$ is a stationary solution of period $\lambda$ and we
evaluate the free-energy ${\cal F}[u_\lambda(x)]$ for the steady
branch $\lambda(A)$, then $(d/d\lambda){\cal F}[u_\lambda(x)] <0$,
i.e. the free-energy decreases with increasing the wavelength.
We are now interested to study the variation of ${\cal F}$ when the
{\it amplitude} of $u_\lambda(x)$ is changed, keeping the period fixed.

Since $u_\lambda(x)$ minimizes ${\cal F}$, it is necessary to go to
the second order. If $u(x)=u_\lambda(x) + \chi(x)$, we get
\bea
\Delta {\cal F} &\equiv& {\cal F}[u_\lambda(x) + \chi(x)] -
{\cal F}[u_\lambda(x)]\\ 
&=&- {1\over 2} \int dx
[ V''(u_\lambda)\chi^2(x) + \chi(x)\chi''(x) ]
\nonumber
\eea

We are interested in fluctuations of the amplitude, so we take
$\chi(x)=\epsilon u_\lambda(x)$ and we ask if $\Delta{\cal F}$
is an increasing or decreasing function of $\epsilon$:
\be
\Delta {\cal F} = {\epsilon^2\over 2} \int dx
[u_\lambda(x)V'(u_\lambda) -u^2_\lambda(x) V''(u_\lambda) ]
\ee

What is relevant is the sign of the integral,
\be
{\cal Q}=\int dx [ u(x)V'(u) -u^2(x) V''(u) ]
\ee
where we have removed the subscript $\lambda$ to lighten notation,
keeping in mind that $u(x)$ is a stationary solution of period $\lambda$.
A positive (negative) ${\cal Q}$ means that $u(x)$ is stable (unstable)
against fluctuations of the amplitude: we want to relate such property
to the behavior of the curve $\lambda(A)$.

Since $V(u)$ is an even function, we can define $\tilde V(u^2)=V(u)$.
Since $V'=2u\tilde V'$ and $V''=2V'+4u^2\tilde V''$, we find
\be
{\cal Q}= -8 \int dx u^4 \tilde V''(u^2) .
\ee

In App.~\ref{app_lambda}, Eqs.~(\ref{eq_dlambda},\ref{eq_dR}),
we show that the sign of $\lambda'(A)$ is related to the sign
of $\tilde V''(u^2)$. For convex or concave potentials $\tilde V(u^2)$,
we can conclude that the sign of $\Delta {\cal F}$ is equal to the
sign of $\lambda'(A)$: if $\lambda'(A)>0$ we have phase instability
(i.e. coarsening), but amplitude stability; if $\lambda'(A)<0$
we have phase stability (i.e. constant wavelength), but amplitude
instability. If the curvature of $\tilde V(u^2)$ changes sign,
it is no more possible to establish such a strict relation between the
signes of $\Delta {\cal F}$ and $\lambda'(A)$ for any value of $A$.

\section{Summary and perspectives}

\subsection{Results}

The two  major results of the present work are (i)  the derivation
of a criterion for coarsening based on the behavior of the
steady state solutions. This criterion holds for several classes of
nonlinear equations that are encountered in various nonequilibrium
systems. The link between the steady state behavior and coarsening
(which is a dynamical feature) has been made possible thanks to the
phase diffusion equation. (ii) The exploitation of the phase
diffusion coefficient has allowed us to derive the coarsening law.
For all known examples which fall within   our classes, we have
captured the exact coarsening exponent. Our analysis has allowed us to make a
quite general statement about the law of coarsening and on the
relevant quantities that are decisive in fixing the coarsening
exponent.

\subsection{Extension to higher dimension}

Usually, an analytical derivation of the coarsening exponent
is made for some equations where their one dimensional
character is essential. While a link between the
phase diffusion coefficient and the behavior of a steady state
branch proves presently difficult to achieve beyond $1d$,
the derivation of the phase diffusion equation can be made at
arbitrary dimension. Our idea according to which $t\sim
\lambda^2/D$ is worth testing in higher dimension.
If it works, since $D$
contains only information on the periodic steady state
solutions, it is sufficient to obtain these solutions to determine
the coarsening law. A numerical  determination of these solutions is
straightforward
and thus the behavior of $D$ as a function of $\lambda$ can easily
be extracted. Then the coarsening law can be obtained without
resorting to a time-dependent simulation.

It must be noted, however, that, for example, in two dimensions
($2d$), there are two independent phase diffusion coefficients.
There are, besides bands, five Bravais lattices in $2d$ and one has
first to determine the steady state solutions, which depend on the
symmetries of the equation. If we take bands as a starting
steady state solution, then there are two principal directions for
the phase evolution: the phase diffusion along the band, and the one
orthogonal to it. Usually, in the vicinity of a threshold (as in the
case of Eq.~(\ref{dispI})) diffusion along the band exhibits a
zig-zag instability, while the one in the orthogonal direction is
associated with the Eckhaus instability~\cite{cross93}. Would only
one of the two coefficients be relevant for the coarsening, or
rather is there a competition between the two directions? It is
clear that if the bands coalesce by keeping their steady state like
symmetry (bands keep their integrity), then, for the CH equation, a
logarithmic coarsening should prevail. We know, however, that the
coarsening law in $2d$ is $\lambda \sim t^{1/3}$~\cite{Bray}. From
this, we expect that both coefficients were essential for the
coarsening. 

One possibility would be to make use of the idea of
diffusion law ($\lambda^2 /D \sim t$) in the  anisotropic case. For
example if we have two diffusion coefficients $D_x$ and $D_y$ along
the two directions, by absorbing these coefficients in $x$ and $y$
coordinates in the phase equation, we arrive to
$(\lambda_x/\sqrt{D_x})^2+ (\lambda_y/\sqrt{D_y})^2 \sim t$, where
$\lambda_x^2+\lambda_y^2=\lambda^2$, and $D_x$ and $D_y$ depend on
$\lambda_x$ and $\lambda_y$.  At long time scales, it is appealing
to expect that isotropy (provided the starting equation enjoys the
rotational symmetry) should be, in principle, restored and
$\lambda_x\sim \lambda_y\sim t^n$ (where $n$ is the
coarsening exponent), and this should complete the extraction of the
scaling with time. 
It is an important task for future investigations to clarify
this point. 

Studies on coarsening
in more than $1d$ may be fairly complicated, even if a lot
of progress has been done in the comprehension of phase
ordering phenomena~\cite{Bray}.
As a matter of fact, in more than $1d$
large scale temporal simulations which are performed with the aim to
 ascertain the coarsening law are extremely time consuming
(even prohibitive in several cases). Note that, as it has been
discussed in Sec.~\ref{sec_n-alpha}, even in $1d$ the cross-over to the
true asymptotic regime may prove to be very long. Thus, if
our idea based on the analysis of phase
diffusion should work in $2d$ (and higher dimension), it
would constitute an important way for the determination of the
coarsening exponent.
We intend to investigate this matter on the GL  equation in $2d$, and
then possibly on other equations.

\subsection{More about the phase equation}

The idea according to which the diffusion coefficient, even in $1d$,
allows one to determine the coarsening law (although it has proven
to be successful for all classes of equations discussed here) calls
for additional comments. Indeed, because the phase equation
$\partial_t\phi=D\partial_{xx}\phi$ exhibits a negative diffusion
coefficient in the coarsening regime, the phase instability leads to
an exponential increase of the phase in the course of time. One
therefore needs to push the expansion in power series of $\epsilon$
to higher orders. It is clear that the next order should lead to the
following linear term $-\kappa
\partial_{xxxx}\phi$ with $\kappa>0$ in order to prevent arbitrarily
small wavelength fluctuations. Nonlinear terms are also needed for
the nonlinear saturation. From general considerations we expect the
first nonlinear term to have the form
$\partial_{x}\phi\partial_{xx}\phi$. This is dictated by the fact
that the phase diffusion equation should enjoy invariance under
the transformation: $x\rightarrow -x$ and
$\phi\rightarrow -\phi$. 

Why and how does the determination of the
coarsening exponent depend solely on the inspection of $D$ is not
completely understood. One possible heuristic  explanation can be
put forward, namely the power counting argument where at large
scales only the first of these terms is decisive. It is of great
importance to clarify this point further, for example by analyzing
the renormalization flow of the nonlinear phase equation at large
time.

\subsection{From coarsening to chaos}

All the evolution equations for which
our general criterion based on $\lambda(A)$
has been derived, exhibit the following
dynamics: (i) they undergo perpetual coarsening, (ii) or they
develop a pattern with a frozen wavelength while the amplitude
increases indefinitely in the course of time. If,
for example, the standard CH equation (\ref{CH}) is modified in the
following manner
\begin{equation}
\partial _t u=-\partial _{xx}[u +\partial _{xx}u -u^3]+\nu
u\partial_x u \label{CH_KS}
\end{equation}
where $\nu$ is a parameter (taken to be positive without
restriction), we obtain a mixture between the CH and KS equations.
For $\nu=0$ we recover the CH equation while for $\nu\rightarrow
\infty$ we obtain the KS limit (upon an appropriate rescaling,
$u\rightarrow u/\nu$). Thus for small enough values of $\nu$ a
coarsening may be expected, while for large values chaos should
prevail. The behavior of this solution was considered by Golovin et
al.~\cite{Golovin}. In the spirit of our analysis we can derive the
phase diffusion equation, or equivalently we can describe the
stability of steady state solutions in a Floquet-Bloch picture. The
branch of steady state solutions in the plane (wavelength-amplitude)
is monotonous up to $\nu=0.47$, beyond which the branch exhibits a
fold singularity. Numerical analysis seems to indicate that the
transition from coarsening to non coarsening occurs at this value.

While it
could not be proven in general that there is a link between the
branch of the steady state solutions and the instability eigenvalues,
it is surprising to see that coarsening stops when the branch
undergoes a fold singularity. On the light of  the various
situations encountered here, it is appealing to speculate  that
whether coarsening occurs or not should be related to considerations
of the behavior of the steady state solutions only. While for the
class of equations like (\ref{eq_cg}) the criterion could be related
to the slope $\lambda(A)$, the criterion may assume a more abstract
form for more general equations. It is hoped to investigate this
matter further in the future.
\appendix

\section{General considerations on $\lambda(A)$}
\label{app_lambda}

Stationary solutions, which play the dominant role in our treatment,
in most cases are determined by a Newton-type equation, \be u_{xx} =
F(u) \ee where the potential $V(u)=-\int du F(u)$ is symmetric and
has a (quadratic) minimum in $u=0$. We are interested in periodic
solutions $u_0(x)$, which correspond to oscillations of period
$\lambda$ and amplitude $A$ within the potential well $V(u)$. We
want to give general criteria to understand if $\lambda$ is an
increasing or a decreasing function of $A$. More precisely, we are
showing how is possible to determine in an easy way the behavior of
$\lambda(A)$ for small and large amplitude $A$. This does not cover
all possibilities, because we may find funny potentials $V(u)$ which
produce an oscillating function $\lambda(A)$: in these cases a
numerical analysis is necessary each time. In all the other cases,
when $\lambda(A)$ has no more than one extremum, the following
analysis applies. At the end of this section we are also
providing an exact expression for $d\lambda/dA$, whatever is the
potential $V(u)$.

It is trivial that in the harmonic approximation, $V(u)\sim u^2$,
$\lambda$ is constant: from a dynamical point of view, this
corresponds to the linear regime. The behavior at small amplitude $A$
is therefore defined by the sign of the quartic term in the potential,
the third one being absent because $V(u)$ is symmetric:
$V(u) \simeq u^2 + a_4 u^4$. It is easy to understand and straightforward
to show (via perturbation theory~\cite{Landau})
that $\partial_A\lambda$
has the sign opposite to $a_4$: $\lambda$ is an increasing (decreasing) 
function of $A$ if $a_4$ is negative (positive).

As a general rule, if the potential is steeper than a parabola, $\lambda$
decreases with increasing the amplitude; in the opposite case,
$\partial_A\lambda >0$. Therefore, if $V(u) \approx |u|^k$ for large $u$,
$\lambda$ increases with $A$ if $k<2$. Using the law of mechanical
similarity~\cite{Landau2}, i.e. scale analysis, we can find that
\be
V(u) \approx |u|^k ~~~ {A\over\lambda^2}\sim A^{k-1} ~~~~
\lambda\sim A^{1-{k\over 2}} .
\ee

If $V(u)$ increases faster than a power law, e.g. exponentially,
$\partial_A\lambda <0$ of course. The same is true if $V(u)$ diverges
at finite amplitude, e.g. $V(u)=1/(1-u^2)$, or if $V(u)$ goes to a finite
value for finite amplitude, but with a diverging force, e.g.
$V(u)=-\sqrt{1-u^2}$.

If $V(u)$ increases slower than a power law, e.g. logarithmically,
or it goes to a constant for infinite $u$, or it has a maximum at
finite amplitude, in all these cases $\lambda$ is an increasing function
of $A$.

Now let us turn to a more rigorous analysis of $\lambda'(A)$.
The exact expression for the period $\lambda$ is
\be
\lambda = 2\sqrt{2} \int_0^A {du\over\sqrt{V(A)-V(u)}} .
\ee
If we use the formula
\bea
&& {d\over dA} \int_0^A du f(u,A) \\
&=& {1\over A} \int_0^A du
[ A\partial_A f(u,A) + u\partial_u f(u,A) + f(u,A) ]
\nonumber
\eea
we get
\be
{d\lambda\over dA} = {\sqrt{2}\over A} \int_0^A du
{R(A)-R(u)\over [V(A)-V(u)]^{3/2} }
\label{eq_dlambda}
\ee
where $R(u)=2V(u)-uV'(u)$.
If, following Sec.~\ref{sec_pot}, we introduce the function
$\tilde V(u^2)=V(u)$, we find that
\be
R'(u) = -4u^3\tilde V''(u^2) .
\label{eq_dR}
\ee
Therefore, if $\tilde V(u^2)$ is a convex (concave) function,
$R(u)$ decreases (increases) with $u$ and
$\lambda$ is a decreasing (increasing) function of the
amplitude $A$.

\section{$\hbox{gCH}$ equation: the sign of $D$}
\label{app_sign-D}

In Sec.~\ref{sec_CH} we found that for the generalized Cahn-Hilliard
equation, the phase diffusion coefficient has the form
$D=q^2\partial_q\langle(\dots)\rangle/\tilde D_2$. In order to establish
the connection between coarsening and sign of $\lambda'(A)$, it is
necessary to show that $\tilde D_2 >0$. 
Let us recall the following:
\bea
&&\tilde D_2 =-\left\langle v{\partial_\phi u_0\over C(u_0)}\right \rangle \\
&&\partial_{\phi\phi} v = {\partial_\phi u_0 \over G(u_0)}\label{eq_v} \\
&&C(u_0)=C(-u_0) > 0 ~~ G(u_0)=G(-u_0) > 0 
\eea

Since $C(u_0)$ is an arbitrary even and positive function and $v$ does not
depend on it, it is necessary to prove that $v\partial_\phi u_0 \le0$,
or, equivalently, that $v\partial_{\phi\phi}v \le 0$.
  From Eq.~(\ref{eq_v}), $v=v_0(\phi) + c_1 + c_2\phi$, where $v_0$
is a $2\pi-$periodic function: we impose $c_2=0$ in order to maintain
this property, while $c_1$ will be fixed later on.

$u_0(\phi)$ is the periodic, bounded trajectory of a particle moving in a
symmetric
potential well, with $V(u)$ being an increasing function between
$u=0$ and $u=A$, the amplitude. $u_0(\phi)$ can be choosen as an odd
function vanishing in $\phi=0,\pm\pi$, with $u_0(\phi)$ and
$-\partial_{\phi\phi}u_0$ having the same sign as $\phi$. If we take the
derivative of $u_0$, use the relation (\ref{eq_v}) and integrate
$\partial_{\phi\phi}v$, we recognize that $\partial_\phi v$ has the same
properties as $u_0(\phi)$. In a similar way, we can say that
$\partial_{\phi\phi}v$ and $\partial_\phi u_0$ are even functions vanishing
in $\pm\phi_0$ (the extrema of $u_0$ and $\partial_\phi v$), that
$\partial_{\phi\phi}v(\phi) >0$ for $|\phi|<\phi_0$ and
$\partial_{\phi\phi}v(\phi) <0$ outside.

Let us now integrate $\partial_\phi v$: $v(\phi)$ is an even
function having maxima in $\pm\pi$ and a minimum in $\phi=0$. 
We fix the constant $c_1$ in such a way that $v(\phi)$ has zeros
in $\pm\phi_0$. Therefore,
since $v(\phi)<0$ for $|\phi|<\phi_0$ and $v(\phi)>0$ outside,
\be
v\,\partial_{\phi\phi}v \le 0 ,
\ee
as we should prove.

\section{The different scenarios for the `crystal-growth' equation}
\label{app_cg}

We consider here Eq.~(\ref{eq_cg})
\be
\partial_t z = -\partial_x\{ B(m)+G(m)\partial_x [C(m)\partial_x m]\}
\ee
where $m=\partial_x z$ and
the functions $B,G,C$ are given by Eq.~(\ref{eq_BGC})
\bea
B(m) = {m\over 1+m^2} , ~~
G(m) = {1+\beta\sqrt{1+m^2}\over (1+\beta)(1+m^2)} ,
\nonumber\\
C(m) = {1+c(1+m^2)(1+2m^2)\over (1+c)(1+m^2)^{3/2} } .
\nonumber
\eea

If we perform a change of variable,
\be
u = \int_0^m ds\, C(s) = {m\over 1+c}{1+c(1+m^2)\over\sqrt{1+m^2}}
\label{eq_u-m}
\ee
we get what we named the {\it full generalized} Cahn-Hilliard equation
\be
\partial_t u = - C(u)\partial_{xx} [ B(u) + G(u) u_{xx} ]
\ee
whose steady states correspond to the trajectories of  a particle moving
in the potential
\be
V(u) = \int du {B(u)\over G(u)} .
\ee

In the general case, it is not possible to invert the function $u(m)$
so as to get an explicit form for $B(u),G(u),C(u)$ and therefore $V(u)$.
However, it will be sufficient to consider the limiting
expressions for $V(u)$ (small and large $u$) in order to discriminate
between the possible dynamical scenarios.

Let us start by considering the case $c=0$, because analytics is simpler.
The function $u(m)$ can be inverted, $m=u/\sqrt{1-u^2}$, showing that
the limit of large amplitude $m$ corresponds for the new variable to
$u\to\pm 1$. The potential $V(u)$ has the form
\be
V(u) =(1+\beta) \int du {u\over\beta+\sqrt{1-u^2}}
\ee
where the parameter $\beta\ge 0$ is easily recognized to be irrelevant,
so we can assume $\beta=0$ and get the approximate potential
$V(u) \sim -\sqrt{1-u^2}$ (see Fig.~\ref{fig_0_0}, inset): 
this case had already been studied by the
authors~\cite{OPL},
finding that $\lambda(A)$ is a decreasing function
(see also the remarks in App.~\ref{app_lambda}).
We can conclude that no coarsening appears if $c=0$.

If $c>0$, $u(m)\simeq cm^2/(1+c)$ for $m\gg 1$ and large amplitude
now means large $u$. In this limit, $m\sim\sqrt{u}$ and
$V(u)\sim (1+\beta)\sqrt{u}/(1+\beta\sqrt{u})$, so that
$V(u)\sim\sqrt{u}$ for $\beta=0$ and $V(u)\sim(1+\beta)/\beta$
for $\beta>0$. In both cases---potential growing as the square root
of the amplitude or going to a constant---$\lambda(A)$ is an increasing
function and there will be asymptotic coarsening. The different behavior
of $V(u)$ for $\beta=0$ and $\beta>0$ suggests different coarsening exponents,
as are actually found.

In the previous paragraph we have considered the limit of large amplitude:
now we are considering the opposite limit of small amplitude. When $m\to 0$
it is trivial to check that $u=m$, but we need the third order correction,
which is easy to calculate from Eq.~(\ref{eq_u-m}): $m \simeq u +(a/2)
u^3$, with $a=(1-c)/(1+c)$. For small $m$,
\bea
{B(m)\over G(m)}&=&{(1+\beta)m\over 1+\beta\sqrt{1+m^2}} \simeq
{(1+\beta)m\over 1+\beta(1+m^2/2)}\nonumber\\
&\simeq & m - {\beta\over 2(1+\beta)}m^3 .
\eea

If now we pass to the variable $u$,
\bea
V(u) &\simeq& \int du \left[\left( u +{a\over 2}u^3\right) -
{\beta\over 2(1+\beta)} u^3 \right]\nonumber\\
&\simeq& {u^2\over 2} + {1\over 8}\left(a-{\beta\over 2(1+\beta)}\right)u^4 ,
\eea
we find that the behavior of the curve $\lambda(A)$ at small $A$ is fixed
by the sign of the quantity $a-\beta/(2(1+\beta))$:
$\lambda$ is an increasing function of $A$ (at small $A$) if that
quantity is negative, i.e. if $a<\beta/(2(1+\beta))$. Using the
expression $a=(1-c)/(1+c)$ we finally find that $\partial_A\lambda>0$
at small $A$, if $c > 1/(1+2\beta)$.

\end{document}